# Structure, preparation, and applications of 2D material-based metal-semiconductor heterostructures

Junyang Tan[1], Shisheng Li[2], Bilu Liu[1,*], and Hui-Ming Cheng[1,3,*]

[1] Shenzhen Geim Graphene Center, Tsinghua-Berkeley Shenzhen Institute and Tsinghua Shenzhen International Graduate School, Tsinghua University, Shenzhen, 518055, P. R. China

[2] International Center for Young Scientists, National Institute for Materials Science, Tsukuba, 305-0044, Japan

[3] Shenyang National Laboratory for Materials Sciences, Institute of Metal Research, Chinese Academy of Sciences, Shenyang, 110016, P. R. China

Correspondence should be addressed to B.L. (bilu.liu@sz.tsinghua.edu.cn) or H.M.C. (hmcheng@sz.tsinghua.edu.cn)

**Abstract**

Two-dimensional (2D) materials family with its many members and different properties has recently drawn great attention. Thanks to their atomic thickness and smooth surface, 2D materials can be constructed into heterostructures or homostructures in the fashion of out-of-plane perpendicular stacking or in-plane lateral stitching, resulting in unexpected physical and chemical properties and applications in many areas. In particular, 2D metal-semiconductor heterostructures or homostructures (MSHSs) which integrate 2D metals and 2D semiconductors, have shown great promise in future integrated electronics and energy-related applications. In this review, MSHSs with different structures and dimensionalities are first introduced, followed by several ways to prepare them. Their applications in electronics and optoelectronics, energy storage and conversion, and their use as platforms to exploit new physics are then discussed. Finally, we give our perspectives about the challenges and future research directions in this emerging field.

**Keywords:** 2D materials, metal-semiconductor heterostructures, structure, preparation, electronics, energy

# 1. Introduction

Since Novoselov, Geim and co-workers isolated graphene from graphite in 2004,[1] the family of two-dimensional (2D) materials has been extended to thousands of members with a variety of electronic properties,[2-3] ranging from insulators (hexagonal boron nitride (h-BN), mica), to semiconductors ($MoS_2$, black phosphorus, $TiO_2$), to metals (graphene). Compared with bulk materials, they exhibit many unique physical and chemical properties, which are related to the electron and phonon confinement effect in the 2D limit.[4] Another advantage of 2D materials is their ability to be integrated into heterostructures. In bulk materials, heterostructures obtained by epitaxial growth play key roles in semiconductor industry, although strict requirements on the matching of lattice symmetry and lattice constant limit the choices of materials.[5] As a result, only a few combinations such as III-V compound-based heterostructures can be epitaxially grown as bulk materials and sophisticated procedures are needed to obtain a sharp interface with minimum strain and few defects. In contrast, because of their atomic thickness, 2D material-based heterostructures with surfaces free of dangling bonds, and weak van der Waals (vdW) forces between adjacent layers, combinations of 2D components are less limited and they can be constructed into lateral heterostructures by in-plane stitching or vertical heterostructures by placing the layers on top of each other.[6-8] In the past few years, many new physical phenomena have been discovered in 2D heterostructures, such as self-similar Hofstadter butterfly states in graphene/h-BN heterostructures[9] and ultrafast charge transfer in $MoS_2/WS_2$ heterostructures.[10] Also, 2D heterostructures based on graphene, h-BN and semiconducting transition metal dichalcogenides (TMDCs) show good performance in electronic and optoelectronic applications, such as field effect transistors (FETs),[11-12] photodetectors[13-14] and light emitting diodes (LEDs).[15-16]

Recently, 2D material-based metal-semiconductor heterostructures or homostructures (denoted MSHSs in this paper) have sparked considerable research interest. Although there are current reports of relatively few MSHSs, most of which focus on graphene-based MSHSs due to limited numbers of 2D metallic materials, increasing numbers of investigations on metallic TMDCs[17] and metal carbides/nitrides (MXenes)[18] provide opportunities for creating new types of MSHSs. MSHSs can integrate the properties of 2D metals and 2D semiconductors in a single structure, and also produce transport[19] and magnetic[20] properties that are absent in the individual components, giving the materials a wide range of properties. For example, MSHSs are seen as candidates for future integrated 2D electronics, in which the metallic components with a high electrical conductivity serve as electrodes and connect with semiconductors with a high-quality interface, leading to a much reduced contact resistance.[19, 21-23] In addition, constructing MSHSs opens a new way of modifying the electronic states and catalytic properties of 2D materials.[24-25] In particular, some metallic 2D materials like $VS_2$ have a high conductivity as well as electrochemical activity,[26] so the heterostructures based on these materials are promising for energy-related applications. As examples of the new physics, some properties of metallic TMDCs, like 2D superconductivity[27] and charge density wave,[28] can be tuned through interlayer coupling in MSHSs. Therefore, MSHSs will become an important topic in 2D materials and more studies are needed for both fundamental research and practical applications.

In this review, we aim to present the latest research progress in the structure, preparation, and applications of 2D material-based MSHSs. A classification based on structure is first given, and we then summarize methods to fabricate both lateral and vertical MSHSs, including direct growth by chemical vapor deposition (CVD), post chemical and phase engineering, and vdW stacking. We also discuss the uses of MSHSs in electronic and optoelectronic devices, energy storage and conversion, and as platforms to explore new physics. Lastly, challenges and future prospects are suggested to motivate more effort in this emerging research area.

**2. Categories of 2D material-based MSHSs**

Due to the versatile compositions of 2D materials and the few requirements to fabricate MSHSs, in theory 2D-based MSHSs comprise a big family with many different combinations. From structural and dimensional points of view, MSHSs are categorized into two types, i.e., all 2D MSHSs and xD/2D hybrid MSHSs (here x is 0, 1, or 3, Figure 1).

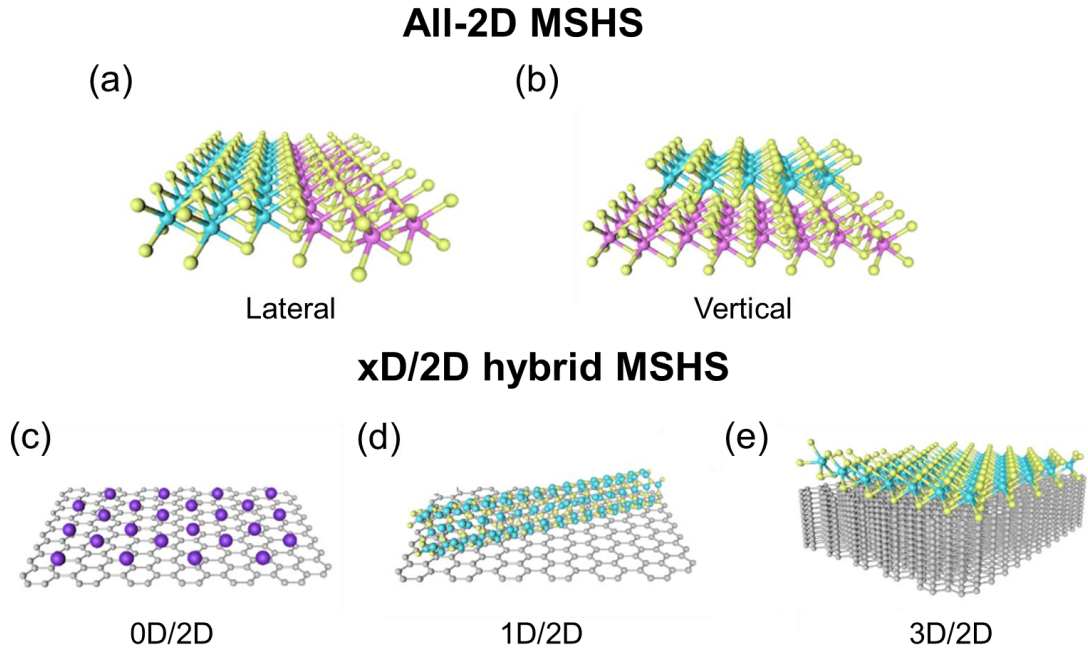

**Figure 1.** Schematics of the different structures of 2D-based MSHSs. All-2D MSHSs include a) lateral and b) vertical structures. Here, xD/2D MSHSs include c) 0D/2D, d) 1D/2D and e) 3D/2D combinations.

**2.1 All 2D MSHSs**

Lateral and vertical heterostructures are the most common configurations for all-2D MSHSs. For lateral MSHSs, 2D semiconductors are stitched with 2D metals within the same atomic plane. In this case, different 2D components are connected with each other by covalent bonds formed at the one-dimensional interface. A sharp interface is the key feature of lateral MSHSs, which may lead to unique optical and electrical transport properties. Due to lattice matching constraints, the lattice constant difference between the metallic and semiconducting components must be considered carefully in order to avoid large stresses or defects occurring at the interface. Limited by current CVD or post treatment preparation approaches, only a few 2D components can be integrated into lateral MSHSs, including graphene-semiconducting TMDC (e.g., graphene-$MoS_2$),[29-30] metallic TMDC-semiconducting TMDC (e.g., $MoTe_2$-$MoS_2$),[31] and MXene-semiconducting TMDC (e.g., $Mo_2C$-$MoS_2$).[32]

Different from lateral MSHSs, 2D metallic flakes and 2D semiconducting flakes can be stacked together with fewer requirements for lattice matching by weak vdW forces, forming vertical MSHSs.[6] Without lattice matching requirements, almost all metallic flakes and semiconducting flakes can be integrated, and the stacking process can be repeated many times to construct superlattices by layer-by-layer stacking.[33] For vertical MSHSs, their electrical, optical and magnetic properties not only depend on the components, but are also influenced by interlayer interactions, and this broadens their applications. In this paper, lateral MSHSs and vertical MSHSs which are made up of component A and B are denoted as A-B and A/B, respectively.

**2.2 xD/2D hybrid MSHSs**

In addition to integrating different 2D materials, integrating them with other dimensional (0D, 1D or 3D) materials can form xD/2D hybrid MSHSs, which expands the existing 2D-based MSHS family. For 0D/2D MSHSs, the most common examples are graphene quantum dots (QDs)[34] or metal nanoparticles[35-36] located on 2D semiconducting TMDCs. Taking the Ag nanoparticle/$MoS_2$ system as an example, the coupling between the two components within the MSHSs can be utilized to tune the photonic behavior of 2D semiconductors by changing light-matter interactions.[35] For 1D/2D MSHSs, a combination of metallic carbon nanotubes (CNTs) and semiconducting TMDCs has been realized by either transfer[37] or CVD growth[38] methods. An FET based on 1D/2D CNT/TMDC MSHSs, where the CNTs and TMDC are respectively used as gate electrodes and the channel material, scales down the lateral size of the heterostructure area to nanometer scale and has a great potential for use in nanoelectronics.[39-40] 3D/2D MSHSs, combining a 2D semiconductor with 3D metallic materials, are very common and have been studied in several devices. Additionally, the integration of graphene and a 3D semiconductor like GaN is another type of 3D/2D MSHS,[41] which shows good stability at elevated temperatures due to the high thermal stability of graphene.

## 3. Preparation of MSHSs

Integrating materials with different properties is an important strategy to build novel platforms for fundamental research and applications, as has been witnessed in the semiconductor industry. In the following section, we will introduce recent progress in the development of preparation methods of lateral MSHSs, including CVD growth and post treatment methods, as well as vertical MSHSs, including vdW stacking, CVD growth and some other methods.

### 3.1 Lateral MSHSs

For lateral MSHSs, due to the chemical bond formed between the metallic and semiconducting components, it is difficult or even impossible to build such structures by aligned transfer or solution assembly methods. CVD growth or similar bottom-up synthesis approaches are the main methods to prepare high-quality 2D materials and have been proven to be able to integrate 2D semiconducting and metallic materials during the growth process to prepare lateral heterostructures. Recently, post treatment has also been shown to be a path to obtain lateral MSHSs by a phase transition or chemical reaction. In the following section, we will discuss the fabrication of lateral MSHSs by CVD growth and post treatment methods.

**3.1.1 CVD growth**

CVD can grow individual 2D materials as well as lateral heterostructures of two different 2D materials. Since the first construction of a graphene/h-BN lateral heterostructure by CVD,[42] a wide range of 2D lateral heterostructures has been produced by this approach.[29-30, 43-53] However, most research has focused on joining different 2D semiconductors, including 2H-2H phase TMDCs like $MoS_2$-$WS_2$,[45-46] $MoSe_2$-$WSe_2$,[47-48] $MoS_2$-$MoSe_2$,[49] $MoS_2$-$WSe_2$,[50-51], 1T'-1T' phase $ReS_2$-$ReSe_2$[52] and even 2H-1T' phase $WS_2$-$ReS_2$.[53]

CVD also offers an effective way to fabricate 2D lateral MSHSs. As one early representative, graphene was chosen as the metallic component to integrate with semiconducting TMDCs, forming graphene-TMDC lateral heterostructures.[54] As shown in Figure 2a, CVD-grown graphene was first transferred onto a $SiO_2$ substrate, followed by patterning by photolithography and oxygen plasma etching. Later, $MoS_2$ nucleated at the exposed edge sites of graphene and expanded into a continuous film filling the bare channels on the substrate, forming graphene/$MoS_2$ lateral MSHSs.[55] Since these exposed edges are random and contain abundant defects, $MoS_2$ prefers to grow along these edges and is polycrystalline, as shown by the false-color dark field transmission electron microscope (TEM) image in Figure 2b. Other methods of growing graphene-TMDC lateral heterostructures have been developed, in which patterning is a critical step for the nucleation of the TMDCs.[29, 56-58] So far, whether covalent bonds are formed between graphene and the TMDC is still unclear. Note that monolayer TMDC consists of three atom layers, in which the metal atom layer is sandwiched between two layers of chalcogens atoms, while the graphene is composed of only one layer of carbon atoms. Therefore, in theory, it is difficult to connect these two components seamlessly by chemical bonds. As shown in Figure 2c, a cross-sectional TEM image of the interface indicates that some TMDC grows into the confined space between the graphene and the growth substrate, forming an overlapping junction rather than an atomically sharp junction.[56] This phenomenon is consistent with the previous report,[55] although the interface seems to be seamless under an optical microscope (OM).

Besides the above-mentioned graphene-TMDC, another type of lateral MSHS is metallic-semiconducting TMDCs. To prepare such a structure, the synthesis of the metallic TMDC is a prerequisite and this is challenging. This is because a high temperature is needed to evaporate the metal precursor, while metallic TMDCs are not usually stable at such temperatures. The recent molten-salt assisted CVD growth process has significantly enhanced the controllability and reproducibility of the growth of metallic TMDCs, by moderating the growth kinetics and reducing the temperature by the addition of salt.[59-60] The choice of growth procedure is also critical to prepare lateral MSHSs, which can be divided into one-step or two-step methods. For a direct one-step CVD method, different 2D components are synthesized and stitched together in a single growth procedure. Although this has been commonly used for synthesizing a MoS$_2$-WS$_2$ lateral heterostructure,[61-62] there is still no report for growing MSHSs by a one-step method. In addition, avoiding the alloying of the two components during growth must be carefully considered.[63] For a typical two-step growth method, the TMDC with the higher synthesis temperature was prepared first. Then, under mild growth conditions, the second TMDC would nucleate and extend from the edge of the first, forming lateral heterostructures. For example, a monolayer 2H-1T' MoS$_2$-MoTe$_2$ lateral heterostructure was synthesized by adjusting the supply sequence of the chalcogens.[31] The scanning transmission electron microscope high-angle annular dark-field (STEM-HAADF) image in Figure 2d shows the atomically sharp interface of the as-prepared heterostructure. Although there is a nonnegligible lattice mismatch of >7% between MoS$_2$ ($b$ = 3.183 Å) and MoTe$_2$ ($b$ = 3.455 Å), the wrinkles in the MoS$_2$ region caused by Te substitution release the strain and avoid the formation of misfit dislocations at the interface. However, as shown in the OM image in Figure 2e, the domain size of 1T'-MoTe$_2$, which would serve as the contact area for the FETs, is still small and needs to be increased. To date, only a few 2D lateral MSHSs have been reported, such as NbS$_2$-WS$_2$[64-65] and VS$_2$-WS$_2$[66]. Unlike what has been reported in semiconducting MoS$_2$-WS$_2$ lateral heterostructures,[45] all these lateral MSHSs have a thicker shell-core structure and a relatively small lateral size, which inevitably hinders their availability for forming contacts. It is believed that synthesizing a range of 2D lateral MSHSs of large size and high-quality will be an important topic for CVD research in the following years. Furthermore, some superior design and experimental setups for growing a 2D lateral semiconductor superlattice, including reverse-flow

CVD[67] and one-pot growth,[68] may offer a potential solution for modifying the growth of MSHSs.

Phase engineering during CVD growth is another way to prepare lateral 2D MSHSs. It is known that for Mo-based and W-based TMDCs, the 2H phase is semiconducting, while the 1T/1T' phase is metallic. Most reported CVD-grown TMDC 1T'-2H homostructures are based on MoTe$_2$,[69-73] due to the small energy difference between its 2H and 1T' phases (35 meV per unit cell).[74] For example, a 1T'-2H MoTe$_2$ lateral MSHS was synthesized using one-step epitaxial growth (Figure 2f-h) and it was found that temperature plays a key role in controlling the phase of MoTe$_2$. At 670 °C, the reaction products are mainly the 2H phase, while the 1T' phase dominates at 710 °C. Therefore, some 1T'/2H MoTe$_2$ mixtures form between 670 °C and 710 °C. Figure 2i shows a homostructure possessing a clean and sharp interface without obvious defects. Tellurizing the pre-deposited Mo oxide film with different Mo/O stoichiometric ratios is another method to obtain 1T'-2H MoTe$_2$ lateral MSHSs with a large size.[75] This method provides a scalable way to integrate 2D material-based channels, electrodes and interconnects into circuits in a single step.

In addition to MoTe$_2$, some other TMDCs like VS$_2$ and TaS$_2$ also have small energy differences between their metallic and semiconducting phases,[17, 76] and the CVD growth of 1T and 2H phases of VS$_2$[77-78] or TaS$_2$[79-80] has been realized. One could obtain lateral MSHSs if both phases were synthesized during one single growth process. However, for some TMDCs like MoS$_2$, it is much more difficult to obtain their corresponding lateral MSHSs because of the large formation energy difference between the metastable metallic and stable semiconducting phases.[81] Specifically, $MoS_2$ tends to form the 2H phase during CVD growth.[82-84] Recently, both a gas–solid reaction[85] and a K ion-assisted CVD method[86] have been reported to produce 1T'-$MoS_2$. However, synthesizing a 1T'-2H $MoS_2$ lateral heterostructure remains challenging. It worth mentioning that a 1T-2H $WS_2$ lateral heterostructure was grown with the aid of synergistic catalysts ($Fe_3O_4$ and NaCl, Figure 2j),[87] but the following photoluminescence and exciton adsorption spectra indicate a direct bandgap semiconducting nature of 1T-$WS_2$, which is inconsistent with present theory.[88] Further investigations on the properties of TMDC polymorphic phases are needed to ascertain their electrical properties.

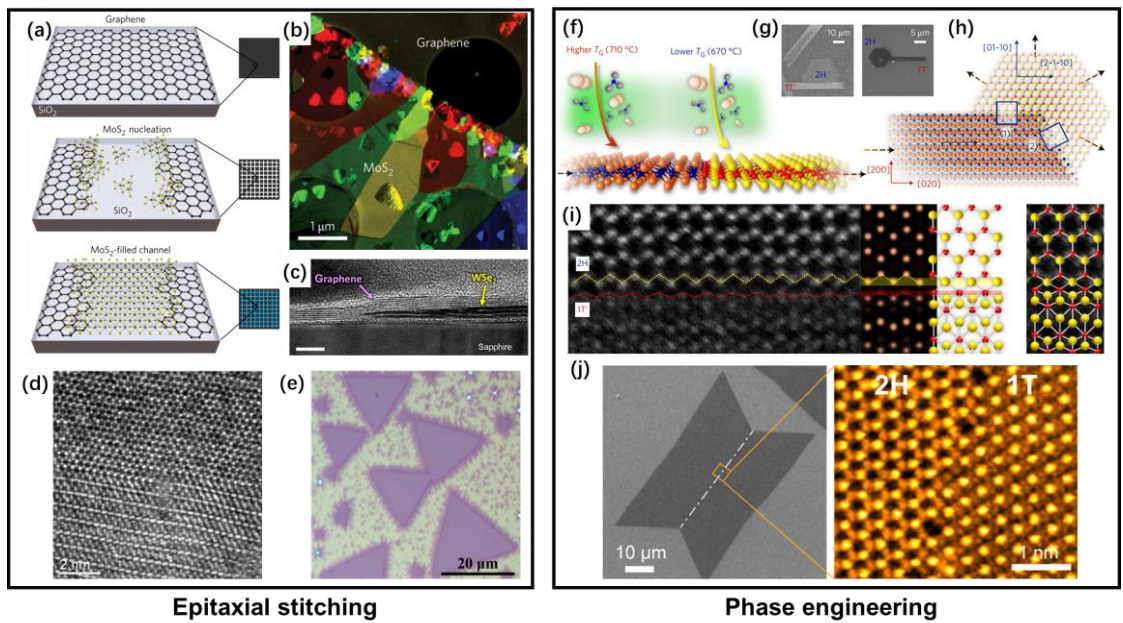

**Figure 2.** Fabrication of 2D lateral MSHSs by CVD growth. a) Schematic of synthesizing graphene-$MoS_2$ lateral MSHSs by CVD. b) False-color TEM image of the interface of graphene-$MoS_2$ lateral MSHSs. a-b) Reproduced with permission.[55] Copyright 2016, Springer Nature. c) Cross-sectional TEM image showing the overlapping junction of a graphene-WSe$_2$ lateral MSHS interface. Reproduced with permission.[56] Copyright 2017, ACS Publications. d) STEM-HAADF image showing the sharp interface of a 2H-1T' MoS$_2$-MoTe$_2$ lateral MSHS. (e) OM image of a 2H-1T' MoS$_2$-MoTe$_2$ lateral MSHS. d-e) Reproduced with permission.[31] Copyright 2017, ACS Publications. f) Schematic of the CVD growth of 1T'-2H MoTe$_2$ by controlling temperature. g) Scanning electron microscope (SEM) image showing the morphology of 1T'-2H MoTe$_2$. h) Schematic of the stitching of the 1T' and 2H phases of MoTe$_2$. i) STEM-HAADF image showing the sharp interface of 1T'-2H MoTe$_2$ (Right: the corresponding atomic model). f-i) Reproduced with permission.[69] Copyright 2017, Springer Nature. j) OM image of a 1T-2H WS$_2$ butterfly and a STEM-HAADF image showing its sharp interface. Reproduced with permission.[87] Copyright 2018, ACS Publications.

### 3.1.2 Post treatment

In addition to the above CVD growth of lateral MSHSs, post treatment is another approach to obtain lateral MSHSs. This involves inducing a phase change in certain areas or chemically modifying them. Both approaches could convert a selected area of a semiconducting TMDC into a metallic one, or the reverse, forming a lateral MSHS. Compared to the complicated in-situ CVD growth of a lateral MSHS, such methods are more convenient and will be discussed in the following sections.

*Post treatment induced phase transitions*

In order to realize a phase transition of TMDCs, ion intercalation, external irradiation, and annealing are commonly used. As early as the 1980s, the ion intercalation of metallic 1T or 1T' phase $MoS_2$ nanosheets was obtained using Li ions during the chemical exfoliation process.[89] The mechanism was explained by the electron transfer from alkali metal to $MoS_2$ increasing the stability of the metallic phase of $MoS_2$, and also significantly decreasing the kinetic energy barrier for the phase transition.[90] By precisely controlling the intercalation process, metallic-semiconducting 1T (or 1T')-2H $MoS_2$[91-92] and 2H-1T $TaS_2$[93] homostructures were prepared by chemically exfoliating the corresponding bulk materials. To enhance the spatial controllability of the ion intercalation process, it could be directly performed on 2D flakes located on a certain substrate (like the CVD samples), combined with patterning techniques like e-beam lithography. Based on this process, 1T-2H $MoS_2$ lateral MSHSs (Figure 3a)[94] and a complicated 2H-1T-2H-1T-2H $WSe_2$ superlattice[95] were made. This lithiation process also depends on the thickness of the TMDC requiring the concentration of the Li reactant to increase with decreasing number of layers. As a result a $MoS_2$ flake with uneven thickness formed homostructures in which the thicker part was converted to the 1T' phase while the monolayer retained the original 2H phase.[96] Apart from alkali metals, zero valent metals can also be intercalated into TMDCs and lead to a change of their electrical properties. For example, Gong et al. intercalated Cu or Co into the vdW gap of bilayer $SnS_2$, converting it from n-type to p-type (Cu-$SnS_2$) and even to a metallic (Co-$SnS_2$) state. Based on this conversion, different patterned $SnS_2$-based MSHSs were obtained with the aid of lithography and, for the first time, 2D metal and n- and p-type semiconductors were stitched together in the same atomic plane as shown in Figure 3b.[97]

External irradiation like an electron beam and plasma can also convert semiconducting 2D materials into those with metallic states. Under the electron beam in the TEM, the irradiated $MoS_2$ region undergoes a phase transition from the hexagonal 2H phase to the metallic octahedral 1T phase, and by spatially moving the electron beam, the 1T-2H interface moved, leading to an expansion of the metallic phase area.[98] Lin et al. directly sculpted 2D semiconductor $MoSe_2$ into metallic MoSe nanoribbons by manipulating a focused electron beam (Figure 3c), which suggested a new way to fabricate metal interconnects for future 2D electronics.[99] A plasma, which is a useful tool for defect engineering[100-101] and reducing the thickness [102-103] of layer materials, can also be used to modulate the phase transition.[104-105] Unlike an electron beam, a plasma is always generated in the whole reaction chamber and cannot achieve selected-area etching.[106] By using a prepatterned mask, a large area of 1T-2H $MoS_2$ MSHSs pattern has been realized under an Ar plasma.[104]

In addition, some experiments have shown that heat treatment leads to the structural rearrangement of TMDCs in order to lower the energy.[107-109] Unlike the recurring transformation between 2H and 1T phases, 2H-$MoTe_2$ shows a unique phase transition under vacuum annealing.[110] As shown in Figure 3d, a parallel bundle of metallic $Mo_6Te_6$ nanowires was formed and stitched to pristine 2H-$MoTe_2$, forming lateral MSHSs.

*Post treatment induced chemical reactions*

MXenes are a promising choice for 2D metals due to their excellent stability and electrical conductivity.[111] Most research has been limited to etching the ceramic MAX phase in solution,[112-113] so the exfoliated MXenes inevitably contain surface functional groups and defects, which degrade their intrinsic properties. Recently, chemical conversion between layer materials and non vdW materials has become a hot topic.[114-115] High-quality MXenes with no surface groups can be obtained by annealing a TMDC in $CH_4$ or $NH_3$, providing a new method for constructing 2D MSHSs.[32, 116-117] For example, Choi et al. patterned a CVD-grown $MoS_2$ film with a mask and in subsequent $CH_4$ treatment the exposed area was converted to metallic $Mo_2C$, while the protected $MoS_2$ area remained unchanged (Figure 3e).[116] The size of lateral MSHS array formed reached hundreds of micrometers. In a similar way, under $NH_3$ treatment, $MoS_2$ was converted to metallic $Mo_5N_6$, and $Mo_5N_6$-$MoS_2$ lateral MSHSs were made by controlling the reaction time.[117] Compared with growth methods like CVD, post treatment such as mentioned above does not need rigorous growth conditions. This topic should be the basis of research that greatly increase the number of existing heterostructure preparation processes.

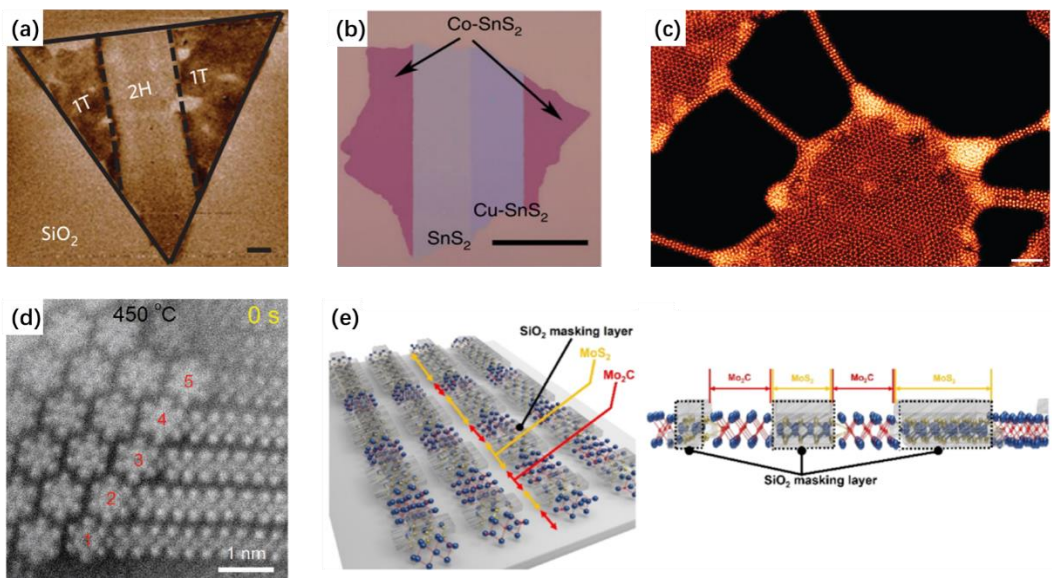

**Figure 3.** Fabrication of 2D lateral MSHSs by post treatment. a) Electrostatic force

microscopy phase image of a 1T-2H-1T $MoS_2$ lateral MSHS formed by Li intercalation. Reproduced with permission.[94] Copyright 2014, Springer Nature. b) OM image of a single MSHS flake containing Co-$SnS_2$, $SnS_2$ and Cu-$SnS_2$ with sharp interfaces. Reproduced with permission.[97] Copyright 2018, Springer Nature. c) STEM-HAADF image showing sculpted metallic MoSe nanoribbons connected to a semiconducting $MoSe_2$ domain. Reproduced with permission.[99] Copyright 2014, Springer Nature. d) Cross-sectional TEM image of a $MoTe_2$-$Mo_6Te_6$ MSHS interface. Reproduced with permission.[110] Copyright 2017, Wiley-VCH. e) Schematic of a $MoS_2$-$Mo_2C$ array formed after $CH_4$ treatment. Reproduced with permission.[116] Copyright 2019, ACS Publications.

### 3.2 Vertical MSHSs

Due to the smooth surface and absence of dangling bonds, isolated 2D materials, e.g. 2D metals and 2D semiconductors, can be directly stacked. CVD is a powerful tool and has the potential to fabricate an array of vertical MSHSs. Besides, there are some other preparation approaches, including solution assembly, wet chemistry synthesis and post treatment methods, which can also stack MSHSs.

### 3.2.1 vdW stacking

Typically, vdW stacking (aligned transfer) consists of two steps. 2D materials are first disassembled from the bulk and are then stacked in the desired sequence by operating a micromanipulator under OM.[118] Since Scotch-tape-based mechanical exfoliation was used to isolate graphene from a graphite crystal,[1] it has become a universal approach to produce a wide range of 2D materials. In the past decade, many efforts have been made to exfoliate high-quality single crystals with a large domain size.[119-121] However, it is difficult to obtain bulk crystals of wafer-scale to exfoliate into 2D flakes. CVD or MBE synthesized 2D continuous films with a few-layer thickness are also a choice as materials for monolayer production. For example, Shim et al. developed a layer-resolved splitting technique based on Ni stickers, that produced a 5 cm-diameter monolayer TMDC on a host wafer from a few-layer sample (Figure 4a).[122] Such an improved exfoliation method is a prerequisite for stacking wafer-scale 2D flakes as MSHSs.

Maintaining clean surfaces during stacking is essential for the integration of 2D materials into high-quality vertical heterostructures. Neither dry nor wet transfer can avoid absorbed contamination or residual polymer because of air, water and amorphous carbon contamination in the ambient transfer environment or organics left when the polymer is dissolved in the solvent. To solve these problems, a programmed vacuum stacking process operated in vacuum conditions was developed by Kang et al., in which wafer-scale TMDCs were stacked with thermal release tape (Figure 4b).[123] The whole transfer process avoids any contact of liquid with the 2D materials being transferred. Cross-sectional STEM measurements (Figure 4c) have verified that the interface of these large heterostructures is contamination-free. Compared to other methods, precise control of the rotation angle is one extra degree of freedom for the manual stacking of 2D materials, and this will provide many more opportunities for vertical MSHSs in twistronics.[124]

It has recently been found that a scanning tunneling microscope (STM) tip can be used to fold graphene from the edge, forming folded bilayer graphene homostructures (Figure 4d-e).[125] This flake-folding operation suggests another way of constructing vertical MSHSs. By picking up the edge of an air-stable 2D semiconductor flake, 2D metallic material could be sandwiched inside the folded semiconductor. This novel method could not only protect the air-sensitive 2D metallic material from ambient conditions, but also control the stacking angle between the top and bottom semiconductor layers and the metal layer between them. Similarly, an atomic force microscope tip or a dome-shaped polymer probe also show potential for this manipulation process.[126]

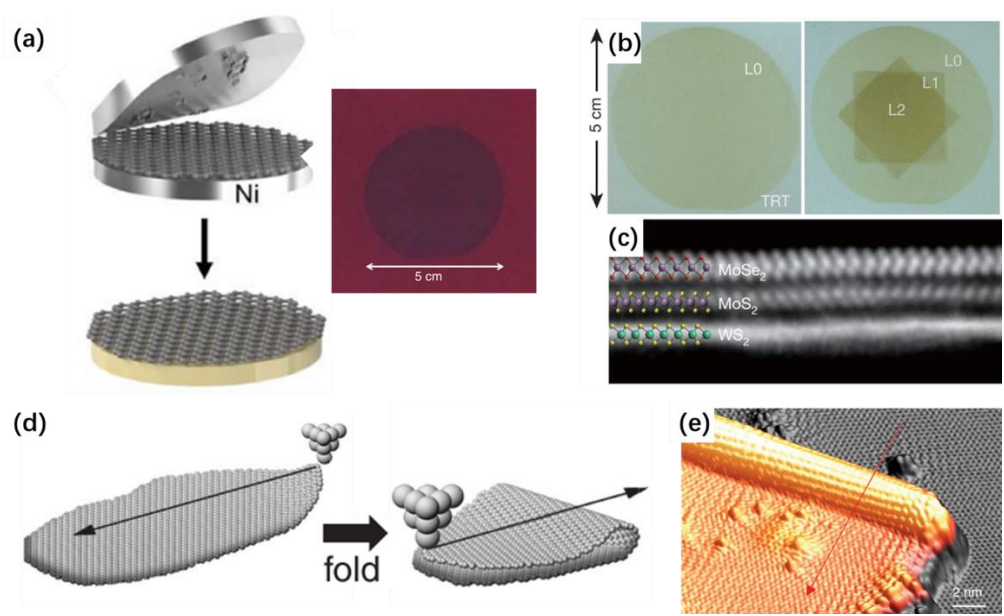

**Figure 4.** Fabrication of vertical MSHSs by vdW stacking. a) Schematic of the layer-resolved splitting technique. (Right: OM image of the isolated monolayer $WS_2$ with a diameter of 5 cm). Reproduced with permission.[122] Copyright 2018, American Association for the Advancement of Science (AAAS). b) OM image of the as-transferred $MoS_2$ films based on the programmed vacuum stacking process. c) Cross-

sectional STEM-HAADF image showing the clean interface of the stacked MoSe$_2$/MoS$_2$/WS$_2$ heterostructure. b-c) Reproduced with permission.[123] Copyright 2017, Springer Nature. d) Schematic of folding graphene into a sandwich structure by an STM tip. e) 3D STM topography of the folded structure. d-e) Reproduced with permission.[125] Copyright 2019, AAAS.

### 3.2.2 CVD growth

Although manual stacking can produce a variety of vertical heterostructures, the process is highly dependent on the researcher's experience and is difficult to scale up. As a bottom-up strategy, CVD method has the potential to prepare a number of vertical MSHSs with high quality and large quantity.

Graphene is an ideal substrate for the epitaxial growth of TMDCs due to their similar hexagonal structure.[127-128] However, the metal substrate (Cu, Ni, etc.) used for graphene growth is not suitable for growing TMDCs due to the reaction between S and these transition metals. Therefore, graphene is typically transferred onto a SiO$_2$/Si substrate for the later growth of TMDCs to form vertical MSHSs.[129] In order to decrease contamination caused by transfer, an all-CVD process is necessary. Pioneering work by Shi et al. produced a two-step CVD method to obtain MoS$_2$/graphene vertical MSHSs.[130] In this work, the liquid precursor (NH$_4$)$_2$MoS$_4$ was deposited on pre-grown graphene, followed by decomposing it into MoS$_2$ at 400 °C (Figure 5a). The epitaxial relationship between MoS$_2$ and graphene was revealed by fast Fourier transform analysis (Figure 5b) extracted from the heterostructure area. Due there being no transfer in the growth process, it produces a clean overlapped area without much contamination (Figure 5c). There are other reports of integrating semiconducting TMDCs with graphene by similar CVD process.[131-132]

Because of their similarity to graphene, semiconducting TMDCs can also be used as templates for the growth of metallic TMDCs on their surface by vdW epitaxy. In such a growth method, a wide range of vertical TMDC MSHSs has been fabricated, including NbS$_2$/MoS$_2$,[133] WTe$_2$/WSe$_2$,[134] NbTe$_2$/WSe$_2$,[135] VTe$_2$/WSe$_2$,[135] TaTe$_2$/WSe$_2$[135] and VSe$_2$/WSe$_2$.[136] Precisely defining the location of the nucleation site on the TMDC 2D flake is critical for the CVD growth of the heterostructure. Recently, defects like boundaries, exposed edges or dislocations in the TMDCs have been found to act as a low energy nucleation sites,[137-140] and based on this, Li et al. recently proposed a universal method for synthesizing 2D vertical MSHSs arrays by creating periodic defect arrays on a large-scale semiconducting WSe$_2$ film, which serve as the nucleation sites for secondary-grown metallic TMDCs (Figure 5d).[141] The periodic arrangement and lateral size of the top 2D metal layer could be modulated by changing the defect pattern. A clean interface is shown by the STEM-HAADF image in Figure 5e.

2D material-based vdW heterostructures have been found to be scrolled into 1D forms. For example, Xiang et al. synthesized a 1D vdW heterostructure by CVD, in which single-wall CNTs were wrapped by outer h-BN and MoS$_2$ nanotubes.[142] This 1D tubular structure provides a new structure design idea for vdW MSHSs.

Apart from these 2D vertical MSHSs, mixed heterostructures including 1D/2D[38,

[143-144] and 3D/2D[145-148] can also be realized by CVD growth. For 1D/2D MSHSs, as demonstrated by Li et al., a CVD method to produce metallic CNT/semiconducting TMDC hybrid films was developed.[38] First, CNTs were grown by a floating catalyst CVD method and transferred onto sapphire. $MoS_2$ then nucleated under the CNTs and continued to grow, forming CNT/$MoS_2$ vertical heterostructures (Figure 5f-g). By oxygen plasma etching, device arrays with $MoS_2$ as channels and $MoS_2$/CNTs as electrodes were fabricated (Figure 5h), and used for photodetection. For 3D/2D heterostructures, $MoS_2$/Au is a classical combination because Au is a good growth substrate for $MoS_2$.[145] Therefore, such 3D/2D MSHSs can be directly obtained by one step CVD.[149] A recent study revealed that during the CVD growth process, an intermediate ($Au_4S_4$) is formed at the interface between Au and the upper $MoS_2$ layer.[148] Compared to transferred 3D/2D MSHSs, this kind of interface reconfiguration shows great potential for tailoring the resistance between the 2D semiconductor and the 3D metal but is little studied, and this may be a novel approach for engineering contacts for future 2D-based electronics.

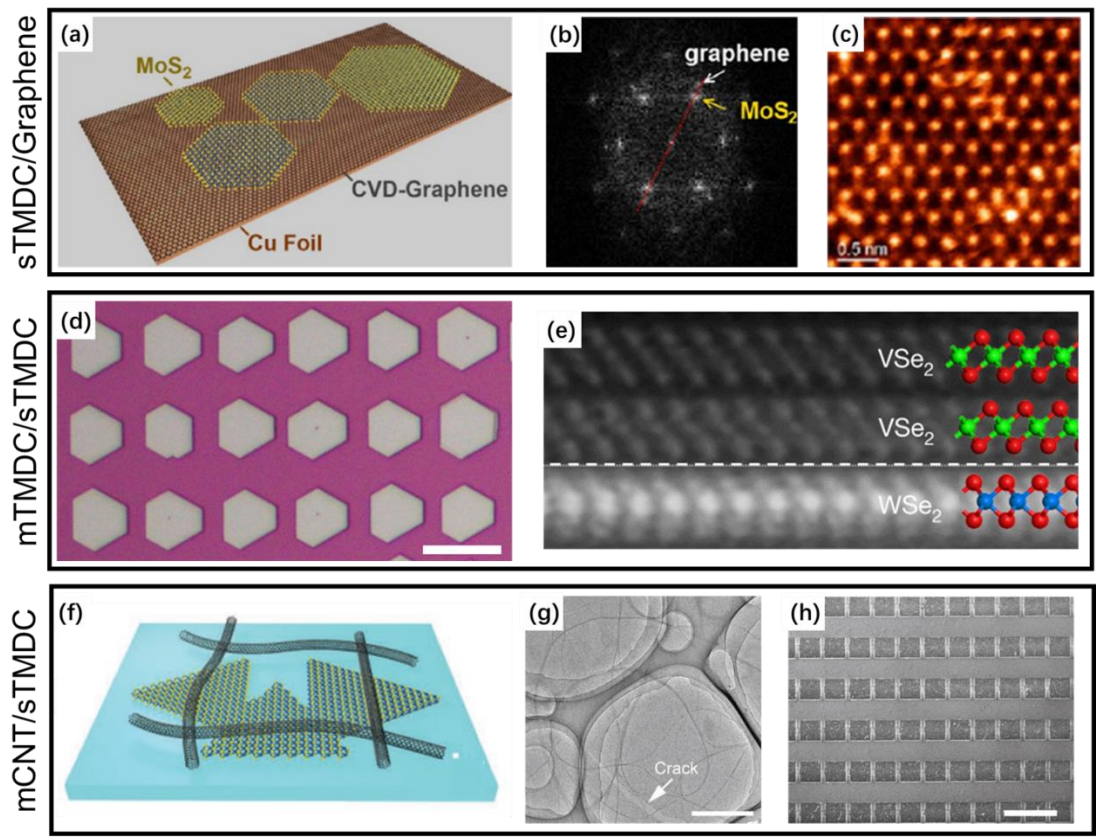

**Figure 5.** Synthesis of vertical MSHSs by CVD growth. a) Schematic of the CVD growth of MoS$_2$/graphene vertical MSHSs. b) Fast Fourier transform image showing the epitaxial relationship between MoS$_2$ and graphene. c) STEM-HAADF image showing the clean surface of a MoS$_2$/graphene vertical MSHS without much contamination. a-c) Reproduced with permission.[130] Copyright 2012, ACS Publications. d) OM image of a VSe$_2$/WSe$_2$ vertical MSHS array. e) Cross-sectional STEM-HAADF image showing the sharp interface of a VSe$_2$/WSe$_2$ vertical MSHS. d-e) Reproduced with permission.[141] Copyright 2020, Springer Nature. f) Schematic of a CNT/MoS$_2$ MSHS film. g) TEM image of CNT/MoS$_2$ hybrid films. h) SEM image of a transistor array based on a CNT/MoS$_2$ MSHS. f-h) Reproduced with permission.[38] Copyright 2018, Wiley-VCH.

**3.2.3 Other methods**

For 2D vertical MSHSs, 2D units are stacked layer by layer with only weak vdW interactions between them. Therefore, with no need to consider constraints like chemical bonding and lattice matching, there are more choices for synthesizing such structures, including solution-assembly and wet chemistry synthesis methods. In addition, post treatment could change the electronic properties of the upper layers while leaving the lower ones in their original state.

*Solution-assembly*

Top down exfoliation is seen as an important strategy for the mass production of 2D materials for industrial applications.[150-151] Recently, some encouraging exfoliation methods have been reported which can produce 2D metallic and semiconducting materials with electronics-grade quality that have few layer or even monolayer thickness.[152-154] Such exfoliated 2D flakes can be dispersed in a variety of solvents, giving dispersions with needed concentrations. Compared with traditional spin coating or vacuum filtration, inkjet printing has shown great advantages in constructing heterostructures.[155] In addition to preparing suitable inks, how to avoid different 2D materials remixing is a key element for constructing vertical heterostructure with sharp interfaces. McManus et al. designed a modified water-based 2D material ink with suitable viscosity and surface tension for inkjet printing.[156] With the help of a suitable binder the 2D materials did not redisperse during the subsequent printing process. Figure 6a shows an as-printed graphene/WS$_2$/graphene MSHS, which has a high photocurrent response and shows potential for use in flexible photosensors. Solution processing also provides a scalable method for building mixed dimensional heterostructures, such as the integration of 0D/2D[34] and 1D/2D configurations.[157] As shown in Figure 6b, after a droplet of graphene QDs was placed on a 2D $MoS_2$ monolayer, they became uniformly dispersed on the surface, which changed the valley polarization of the underlying $MoS_2$. Limited by the lateral size and thickness of present exfoliation technique, it is difficult for solution assembled MSHS-based devices to match the performance of samples produced using non-scalable methods. However, due to its high yield and low cost, solution-assembly must be an important process for future industry-scale flexible electronics.

*Wet chemistry synthesis*

Hydrothermal and solvothermal processes are widely used to synthesize nanomaterials, including 2D TMDCs.[3] The reaction always happens in a sealed environment with high temperature and high pressure, and water or organic solvents act as reaction media. By adjusting the precursors, this simple and inexpensive method has been used to obtain a series of vertical MSHSs, including $rGO/MoS_2$,[158] $MoS_2/CoSe_2$,[159] $Sn_{0.5}W_{0.5}S_2/SnS_2$[160] etc. Another advantage of these processes is the various complex structures of the synthesized materials, which could contribute to their electrocatalysis and energy storage performance.

A hot-injection reaction, in which a precursor solution is rapidly injected into the reaction solvent, is another method to synthesize MSHSs.[161-162] Recently, Sun et al. proposed a universal approach to decorate noble metals (Au or Ag) on W- or Mo-based TMDC nanosheets to form 0D/2D and 2D/2D hybrid structures.[162] As shown in Figure 6c, Au tends to deposit as nanoparticles distributed on all $MS_2$, $MSe_2$ and $MTe_2$ compounds (M=Mo or W), while when Ag spreads over $MTe_2$, forming a Ag/$MTe_2$ 2D/2D MSHS. These different deposited morphologies are explained by the different reducing ability of the TMDCs.

*Post treatment*

Shining an energy beam like a laser on 2D multilayers may cause the formation of vacancies or a phase transition, which may change the electronic states of the upper layers and form vertical homostructures. An interesting study by Cho et al. shows that laser-irradiation converts the upper semiconducting 2H $MoTe_2$ layers into a metallic 1T' phase with a reduction of thickness.[163] Similarly, for 2D $PdSe_2$, which is a novel semiconductor with a layer-dependent bandgap, external irradiation was found to induce a semiconducting-to-metallic phase change, leading to the formation of $PdSe_2$/$Pd_5Se_{17}$ vertical MSHSs (Figure 6d).[164] The generated Se vacancies play a key role in the structural change of $PdSe_2$.[165] These examples have shown the ability of post treatments in preparing vertical MSHSs. Such direct processing of 2D materials is scalable and appropriate for the potential batch fabrication of MSHSs.

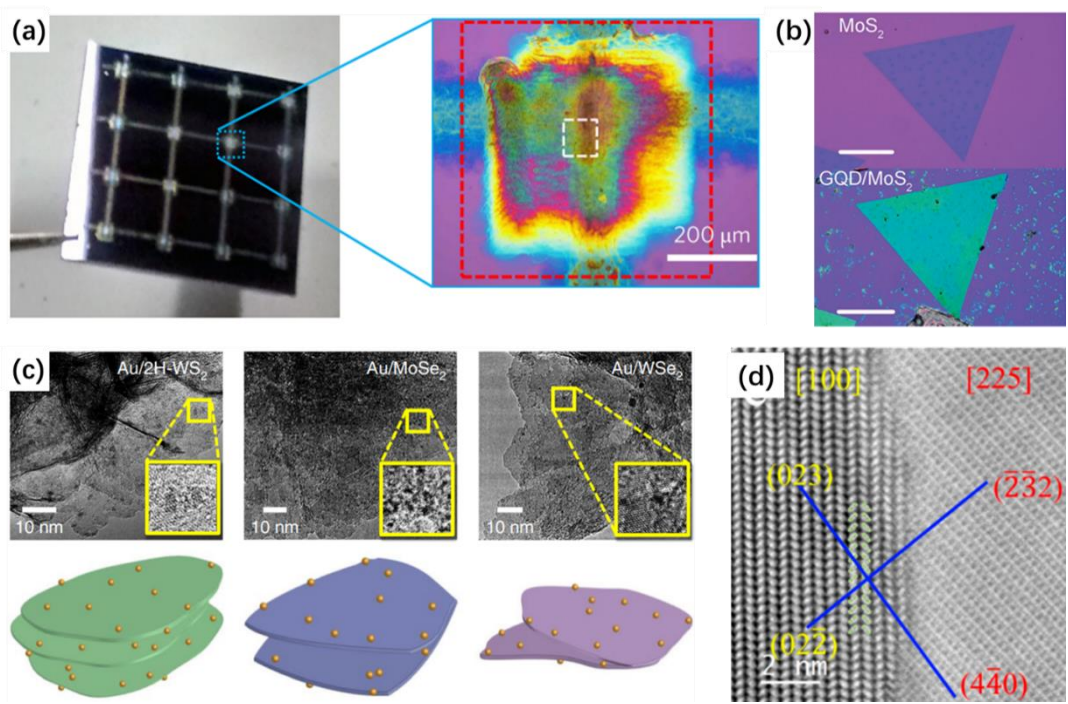

**Figure 6.** Synthesis of vertical MSHSs by solution-assembly, wet chemistry synthesis and post treatment. a) Optical photo of an as-printed graphene/WS$_2$/graphene MSHSs. (Right: OM image of one of the MSHSs). Reproduced with permission.[156] Copyright 2017, Springer Nature. b) OM image of pristine MoS$_2$ and a graphene QD/MoS$_2$ MSHS. Reproduced with permission.[34] Copyright 2015, Wiley-VCH. c) TEM image of Au nanoparticles decorated on TMDC flakes and the corresponding models. Reproduced with permission.[162] Copyright 2020, Springer Nature. d) Cross-sectional STEM image showing the sharp interface between PdSe$_2$ and Pd$_{17}$Se$_{15}$. Reproduced with permission.[164] Copyright 2019, ACS Publications.

## 4. Applications of 2D MSHSs

With continued research in MSHSs, this integrated metal/semiconductor structure has shown potential for use in electronic and optoelectronic devices, energy storage, and electrocatalysis, and as a platform to explore new physics like 2D magnetism and superconductivity. In this section, recent progress on the application of MSHSs in these areas is discussed.

**4.1 Electronic and optoelectronic devices**

2D materials are seen as promising channel materials in transistors.[166] On the one hand, their natural atomic thickness gives 2D semiconductors short conduction channels, leading to a rapid switching performance, and their thinness ensures that they have the good flexibility required for wearable devices. Although in recent years, great efforts have been made to obtain 2D-based transistors for electronic devices, there is still great difficulty in bringing this to fruition and there are gaps between theory and practice. Taking $MoS_2$ as example, which is a representative 2D semiconductor, it has a theoretical room-temperature electron mobility of over 400 $cm^2$ $V^{-1}$ $s^{-1}$,[167] but experimentally it is usually below 200 $cm^2$ $V^{-1}$ $s^{-1}$.[168-169] Apart from defects in the 2D channel, the large resistance caused by the less-than-ideal contact between the channel and the electrode is mainly responsible for the reduced value.[21, 170-171] Therefore, optimizing contact is of significance for high performance devices.

The Schottky barrier height (SBH) is a measure of the quality of a metal-semiconductor junction, where a higher SBH means a higher energy barrier for carriers to overcome when crossing the junction. Ideally, the SBH could be described by the energy difference between the work functions of the metal and the band edge of the semiconductor, namely the Schottky-Mott rule.[172-173] This means that the contact could be modified by selecting a different metal with a suitable work function depending on the channel material. Unfortunately, for a typical top contact, the inevitable Fermi level pinning effect (FLPE) means that the selection of a different metal may have a limited role. It may be attributed to the poor interface states between the metal electrode and the semiconductor produced by conventional device fabrication processes, such as metal deposition and lithography.[174-176] As shown in Figure 7a, obvious defects, fractures and an amorphous area appear at the interface during "high energy" deposition,[177] which causes a nonnegligible FLPE and leads to the poor transport performance of the FETs.

One major advantage of MSHSs is the low contact resistance between the electrode and the channel. For vertical MSHSs, a 2D metallic material has face-to-face contact with a 2D semiconductor with only weak vdW forces between them. This novel contact geometry is efficient in eliminating the FLPE. Liu et al showed the origin of the weak FLPE of vdW contact by theoretical simulation.[178] Taking 1T-$MoS_2$/1H-$MoS_2$ as an example, the distribution of the states within the bandgap of the interface is almost dominated by the metal, which means the combination does not give rise to metal-induced gap states (MIGS) so that the FLPE does not occur. Experimentally, with the continuous discovery of 2D metallic materials, the family of vdW contact bi-materials is expanding by either transfer methods or epitaxial growth. Graphene is the first choice for a 2D electrode. In an early study, Das et al. assembled all 2D thin film transistor with graphene as the electrode, $WSe_2$ as the channel, and h-BN as the dielectric material. This low-resistance device exhibited a high $I_{ON/OFF}$ ratio of $10^7$ as well as a high carrier mobility of 45 $cm^2$ $V^{-1}$ $s^{-1}$.[179] Metallic TMDCs have also become promising electrode candidates. Ji et al. synthesized a few-layer metallic $VS_2$ nanosheet by CVD and stacked it on MoS$_2$ as the electrode. Compared to a Ni/Au-MoS$_2$ FET, the contact resistance of the VS$_2$/MoS$_2$ FET was decreased more than 75%.[77]. Although recently reported transfer[177] or imprint[180] 3D electrode methods could also form defect-free 3D/2D vdW contacts (Figure 7b), they require either a precise transfer process or not too high a temperature of combination, which is difficult to match with current electronic fabrication requirements.

For lateral MSHSs, a 2D semiconductor and a 2D metal have been seamlessly stitched together with a sharp interface, generating an edge contact. Analogous to vdW contact, the FLPE should also be negligible for edge contact geometry, but the underlying mechanism is different.[19, 181-182] As ab initio simulation based on a monolayer 1T-2H MoS$_2$ lateral MSHS shows that MIGS will penetrate the Schottky barrier, significantly reducing the energy gap for carriers to move from the metal to the semiconductor.[182] Kappera et al. observed superior transport performance in such lateral MSHSs. A 1T-2H MoS$_2$ lateral heterostructure was prepared by Li ion intercalation, and contact between the metallic 1T phase and the semiconducting 2H phase was ohmic with a low value of 200–300 Ω μm at zero gate bias (Figure 7c), which is less than a fifth of typical Au top contact.[94] This Li ion intercalation strategy also works for other 2D materials like WSe$_2$.[95]

As discussed above, for both vertical MSHSs with vdW contact and lateral MSHSs with edge contact, the FLPE is negligible, meaning the Schottky-Mott rule should be followed for the selection of the individual MSHS components. In order to lower the SBH, for n-type semiconductors 2D metals with low work functions will match better, while 2D metals with a high work function are preferred for p-type ones. Figure 7d shows a library of simulated work function values for different 2D metallic materials and the bandgap of 2D semiconductors.[178] It is obvious that 2D metallic materials have a wider range of work functions than traditional 3D bulk metals, from the low work function of highly N-doped graphene (as low as Sc) to the very high value of 2H-NbS$_2$ (even much higher than Pt), which provides a really large range of ways to moderate the SBH of MSHSs. Recently, Zhang et al. prepared VSe$_2$/MoSe$_2$ and VSe$_2$/WSe$_2$ vertical heterostructures and used a Kelvin probe force microscope to investigate the Fermi-level difference between the metallic VSe$_2$ and the semiconducting components.[136] For n-type MoSe$_2$, there is Ohmic contact at the interface of VSe$_2$/MoSe$_2$ heterostructure, while for p-type WSe$_2$, VSe$_2$ tends to combine with it to form a Schottky barrier diode. In conclusion, the Schottky-Mott rule will provide a guideline for future MSHS design.

For the commercial production and application of 2D electronics, 2D material-based MSHSs should be able to be used in the device manufacturing process. The CVD production method, in which a metallic electrode is bridged on a semiconductor during in-situ growth without any destructive post-treatment processes, will be a possible choice. Ideally, because of its high controllability CVD should be able to produce a patterned array of MSHSs, which is a precondition for integrated circuits. As mentioned earlier, Li et al. synthesized large-scale WSe$_2$ transistors arrays (Figure 7e), with superimposed VSe$_2$ working as source and drain electrodes.[141] Thanks to the high-quality interface of the MSHSs, the as-fabricated WSe$_2$ transistors showed Ohmic contact (Figure 7f) with a cumulative mobility of around 100 cm$^2$ V$^{-1}$ s$^{-1}$, while the contact performance of the WSe$_2$ with deposited Cr/Au electrodes was much poorer with a mobility of around 10 cm$^2$ V$^{-1}$ s$^{-1}$ and a nearly 1000 times lower $I_{ON/OFF}$ ratio.

Besides electronics, optoelectronics is another important application for 2D MSHSs. One apparent advantage of 2D-based MSHSs is their nm-scale thickness, ensuring tunable light absorption and transmittance, which is crucial for optoelectronic devices such as photodetectors and photovoltaic devices. For MSHSs, in addition to their superior contact performance, the metallic layer could help in the separation of photogenic carrier pairs, leading to a high photoresponse efficiency. For example, Yu et al. fabricated a graphene/MoS$_2$/graphene vertical MSHS photodetector with a maximum external quantum efficiency of 55% and an internal quantum efficiency up to 85% ($\lambda$=488 nm).[183] This sandwich structure provides a new approach to moderate photocarrier generation, separation and transport with the help of an external back-gate voltage. In addition, a metallic layer could also be inserted in the p/n junction as an interlayer, forming a p-type/metal/n-type vdW heterostructure. In another study, Li et al. produced a MoTe$_2$/graphene/SnS$_2$ MSHS and investigated its photodetecting performance.[184] For this vertical heterostructure, MoTe$_2$ is in the 2H-phase and is p-type, while SnS$_2$ is an n-type semiconductor. The sandwiched graphene not only accelerated electron transport but also delayed interface charge traps, leading to a much higher photoresponse. The photodetector achieved an ultrahigh response of over 2600 A W$^{-1}$ and a specific detectivity up to $1.1 \times 10^{13}$ Jones over a broad spectrum from ultraviolet to short-wave infrared. Until now, for MSHS-based optoelectronics, most reports have used graphene as a conductive layer rather than 2D metal because it has a low theoretical absorption value of 2.3% per layer.[185] The advantages of other metallic materials in optoelectronic applications still need to be explored.

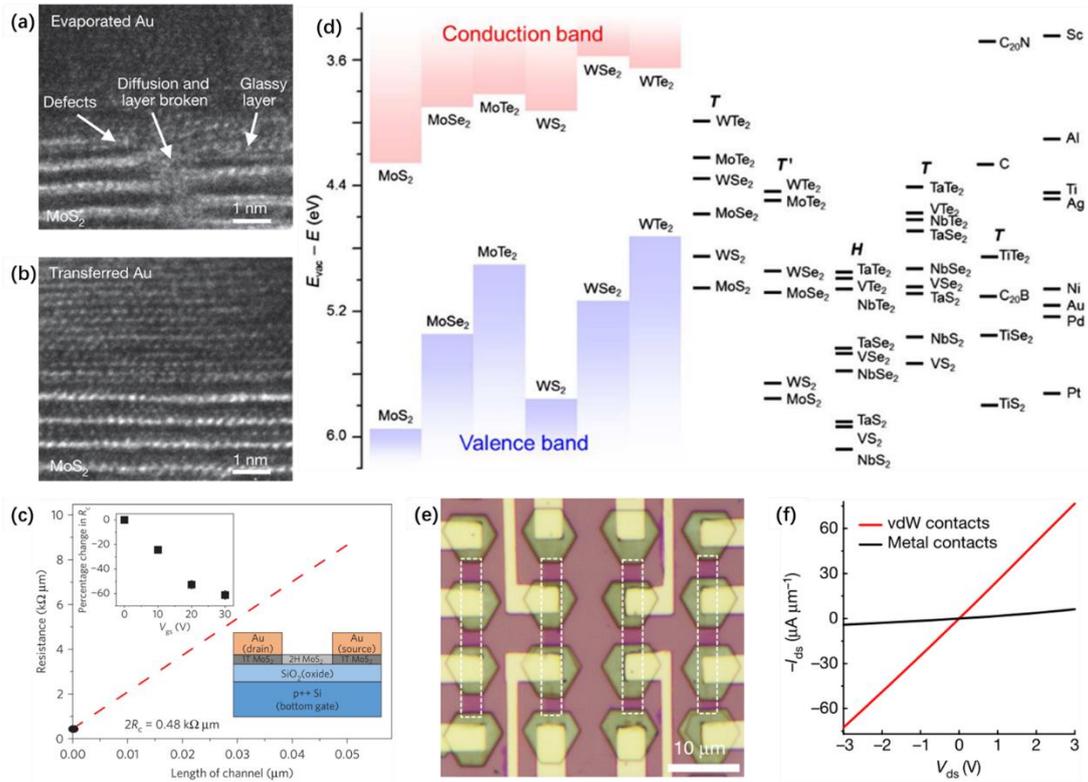

**Figure 7.** MSHS electronics. a,b) Cross-sectional TEM image of the interface of (a) conventional evaporated Au and (b) transferred Au on MoS$_2$. a,b) Reproduced with permission.[177] Copyright 2018, Springer Nature. c) Resistance versus 2H-MoS$_2$ channel lengths for a 1T-2H MoS$_2$ lateral heterostructure. (Left inset: the percentage decrease in contact resistance with gate bias; Right inset: model image of the device). Reproduced with permission.[94] Copyright 2014, Springer Nature. d) Band alignment between 2D metallic and semiconducting materials. (Left: conduction band and valence band of 2D semiconducting materials; Right: Work function of 2D metallic materials.) Reproduced with permission.[178] Copyright 2016, AAAS. e) OM image of a transistor array based on VSe$_2$/WSe$_2$ vertical MSHSs. f) Comparisons of transistors with synthetic vdW contacts and conventional top contacts formed by directly depositing the metal electrode. e,f) Reproduced with permission.[141] Copyright 2020, Springer Nature.

## 4.2 Energy storage and conversion

Due to the increasing demand for energy and the resulting environmental problems, developing environmentally-friendly and sustainable energy pathways with high efficiency is becoming important. Recently, many 2D materials have shown excellent performance as electrodes or electrocatalysts for water splitting, including graphene, TMDCs (like $MoS_2$), MXenes (like $TiC_2$), etc.[186-188] In addition to defect and phase engineering, which are common methods to improve the intrinsic performance of 2D materials,[189-190] integrating them with different components into heterostructures provides a novel way to change the performance by synergetic chemical coupling effects.

Ion batteries are important energy storage systems. Thanks to their layer structure and large specific surface area, 2D material-based electrodes have abundant ion intercalation sites, leading to a high charge storage ability. Stacking different 2D materials into heterostructures will also modify the battery performance.[25] Compared with pristine 2D building blocks, vdW heterostructures have a weaker interaction between the layers, which increases the interlayer spacing between them.[191] This allows the accommodation of more ions and increases the ion diffusion speed, so both energy and power densities are improved for vdW heterostructures. Especially for MSHSs, in which metallic materials increase the electric conductivity of the overall electrode, it ensures superior performance during high-speed charge-discharge cycling. In addition to graphene-based MSHSs, which have become a common electrode material,[158, 191-192] some combinations of 2D semiconductors and other metallic materials (metallic CNTs, TMDCs and MXenes) may also be useful.[193-195] For example, Chen et al. synthesized $MoS_2$/MXene MSHSs by the in-situ annealing of hybrids of $Mo_2TiC_2T_x$ and sulfur particles.[196] The prepared compound has a high reversible average capacity of 548 mA h g$^{-1}$ at a current density of 50 mA g$^{-1}$ for Li-ion batteries. It also has a much higher capacity and better cycling stability than individual Mo$_2$TiC$_2$T$_x$ and MoS$_2$.

For electrocatalysis, water electrolysis has become a crucial technology for energy conversion. Through the hydrogen evolution reaction (HER), H$^+$ or water is reduced at the electrode/electrolyte interface, generating H$_2$. MoS$_2$, which is an important HER catalyst because its edges and intrinsic defects function as active sites.[197] Combining MoS$_2$ and graphene in MSHSs is an effective way of increasing the HER activity, which can be explained by both electronic coupling and an overall increased conductivity.[198-199] Also, for MSHSs catalysts, metallic TMDCs may replace the function of graphene because they have a much higher performance originating from their basal-plane active sites.[200] Some TMDC-based MSHSs, such as 1T-MoS$_2$/2H-MoS$_2$,[201] MoSe$_2$/NiSe,[161] MoS$_2$/VS$_2$,[202] MoS$_2$/CoSe$_2$[159] have been synthesized and show a better HER performance than the single components. For these MSHS catalysts, the underlying mechanism needs more research. Some studies attribute the better performance to synergistic effect; however, more detailed understanding is very important to design suitable MSHS catalysts with superior performance.

**4.2.3 Platform to exploit new physics.**

In addition to the above-mentioned applications, 2D MSHSs are also a good platform to explore new physics. Metallic TMDCs have attracted a lot of attention because of their unique physical properties, including 2D superconductivity,[27] magnetism[203] and charge density wave (CDW)[28]. How to take advantage of and change the new physics is always an important research topic. Until now, many methods have been explored, including thinning, straining, and hole/electron doping the 2D materials.[204-206] Stacking a 2D semiconductor with a metallic TMDC provides an effective way of changing these properties. Unfortunately, due to the difficulty of mechanically exfoliating or synthesizing 2D metallic TMDCs, it is difficult to obtain their corresponding heterostructures. Although only a few pioneering studies have been reported, it is believed that metallic TMDC-based MSHSs are a suitable platform for the fundamental study of these 2D-related new physics.[207-209]

Wang et al. transformed the top layer of bulk 1T-$TaS_2$ into the 2H phase by annealing, forming a monolayer 2H-$TaS_2$/bulk 1T-$TaS_2$ vertical homostructure.[207] The superconducting transition temperature ($T_c$) of the sample is around 2.1K, which is about a threefold increase over that of bulk 2H-$TaS_2$. The authors claimed that the suppressed $3 \times 3$ CDW and charge doping from the substrate contribute to the higher $T_c$. This work also provided a way to investigate the proximity effect between 2D CDW and superconductivity. Later, the same group changed the CDW order in 1T-$TaS_2$ by forming a 1T-$TaS_2$/black phosphorus vdW heterostructure.[209] A nearly commensurate CDW (NCCDW) exists until 4.5K, much lower than the phase transition temperature from NCCDW to CCDW. Interestingly, due to the tuning of anisotropic black phosphorus substrate, the transport properties of 1T-$TaS_2$/black phosphorus are also anisotropic.

Stacked vdW metal-semiconductor heterostructures are an important platform for studying and understanding the 2D physics, and the twist angle is an additional important factor.[210-211] For stacked MSHSs, twisted structures are still unreported and there are many opportunities to explore the relationship between the interlayer interaction and the twist angle.

## 5. Conclusions and Outlook

We have reviewed recent progress in 2D material-based MSHSs. From the structural viewpoint, MSHSs are classified into all 2D MSHSs (lateral and vertical) as well as xD/2D hybrid MSHSs (0D/2D, 1D/2D, and 3D/2D). Several preparation methods for MSHSs have been introduced and comparisons between these methods are discussed. Finally, the uses of MSHSs in electronics, optoelectronics, energy storage and conversion, and fundamental studies have been considered. MSHSs are still in their infancy and there are many issues waiting to be investigated, such as finding new 2D metallic materials, improving control of their preparation, understanding the interfaces, and exploring further applications, as illustrated in Figure 8.

First, new preparation methods for 2D metallic materials need to be developed. Compared to 2D semiconductors, the number of 2D metals is very limited. Besides graphene, metallic TMDCs and MXenes, some newly emerging 2D metals, such as metallene,[212-214] borophene,[215-216] $Fe_3GeTe_2$,[217] have become potential components to expand the MSHS family. However, they are not easy to prepare and may be unstable. For metallene and borophene, there are no natural parent layer materials for exfoliation, and only bottom-up strategies with rigorous conditions can synthesize these 2D materials. For materials like $Fe_3GeTe_2$, bulk layer materials can be synthesized by the flux method, but the strong interaction between the layers makes it difficult cleave them into the 2D form. Therefore, using a special substrate with which they have a strong interaction or intercalating specific ions/molecules into bulk layer materials may provide a new way to prepare the 2D form. In addition, some recently developed new methods, for example, the dissolution-precipitation growth method, may be helpful to grow such 2D materials and need further development. In addition, many metallic materials easily degrade in ambient conditions and how to protect them during the preparation process is an important topic. To this end, Briggs et al. synthesized 2D Ga, In and Sn in the confined space between a silicon carbide substrate and graphene, which significantly improved their stability.[218] Such in situ encapsulation may be a way to grow unstable 2D materials, especially metallic ones. The use of stable 2D materials like graphene oxide or TMDCs as substrates to grow and stabilized unstable or meta-stable 2D materials is another interest direction needs investigations, which is also a potentially important platform to discover new materials.

Second, the controllability of preparing MSHSs needs to be improved. CVD is most likely to achieve scalable of electronic-grade 2D MSHSs for the semiconductor industry,[219] however, compared to growing individual 2D materials, controlling the growth of heterostructures is far more difficult. In a conventional CVD process for preparing MSHSs, it is difficult to control the nucleation sites for the second component located at the edges or on the surface of the first one, which leads to a low yield. In addition to creating defects on TMDCs to serve as nucleation sites,[141] coating with seeding promoters[83] or depositing Au seeds[220] at selected positions during the time between the two growth steps may also guide the nucleation process. We believe that these methods will also contribute to the aligned or patterned growth of MSHSs. For MSHSs obtained by post treatment methods, although patterning techniques like e-beam lithography have been used to enhance the controllability, the patterning process is complicated and may leave residues on the surface. Techniques with high spatial resolution, such as electron beam irradiation, may help achieve the lithography-free preparation of MSHSs. Recently, substitutional doping has also become a strategy to increase the electrical conductivity of semiconducting TMDCs,[22-23] even to producing metal-like conduction[221]. To realize this goal, a universal doping strategy with tunable dopant concentrations is needed,[222] and the influence of different dopant atoms remains to be determined. Beyond simple MSHSs with two components, the design and fabrication of complicated MSHSs or super-lattices with multiple alternatively stitched or stacked components is an interesting topic to pursue.

Third, more effort is needed to study the new physics in MSHSs which is quite different from that of their single components. Researchers have paid much attention to the electronic transport properties in MSHSs due to their potential use in 2D electronics. The state of the interface plays a key role in determining the carrier transport behavior across the interface, and defects or contaminations at the interface will significantly degrade the device performance. For lateral MSHSs, there is accumulated strain at the 1D interface because of the different lattice constants. However, the relationship between strain and the change of transport behavior between a metal and a semiconductor are still not clear. Recently, Ugeda et al. observed topologically protected states at the interface of 1T'-2H WTe$_2$,[223] which will broaden the study of the interface of lateral MSHSs. For vertical MSHSs, in addition to transport properties, some pioneering studies have investigated or predicted their optical and magnetic properties induced by the interlayer coupling.[20, 208] We believe that this will be an important topic for 2D physics.

Finally, in addition to the applications mentioned in Section 4, more applications need to be explored based on the intriguing coupling between metals and semiconductors in MSHSs. For example, some pioneering studies have shown the suitability of 2D material-based MSHSs in the sensing of gases,[224-225] organics,[226] and DNA.[227] In addition, the ultrafast excitonic behavior in MSHSs could broaden the applications of 2D materials in optoelectronic devices,[208] and the unique spintronic states in Fe$_3$GeTe$_2$-based MSHSs will be used in future magnetic storage applications. Meanwhile, some 2D materials are anisotropic in electronic and optical properties, and considering the large intrinsic anisotropic of 1D system, the 1D/2D MSHSs may be particularly interesting in polarized optics. Along with the development of advanced fabrication techniques and a deep understanding of the physics involved, it is believed that 2D material-based MSHSs will offer many uses in materials science, electronics, energy, physics, and nanotechnology.

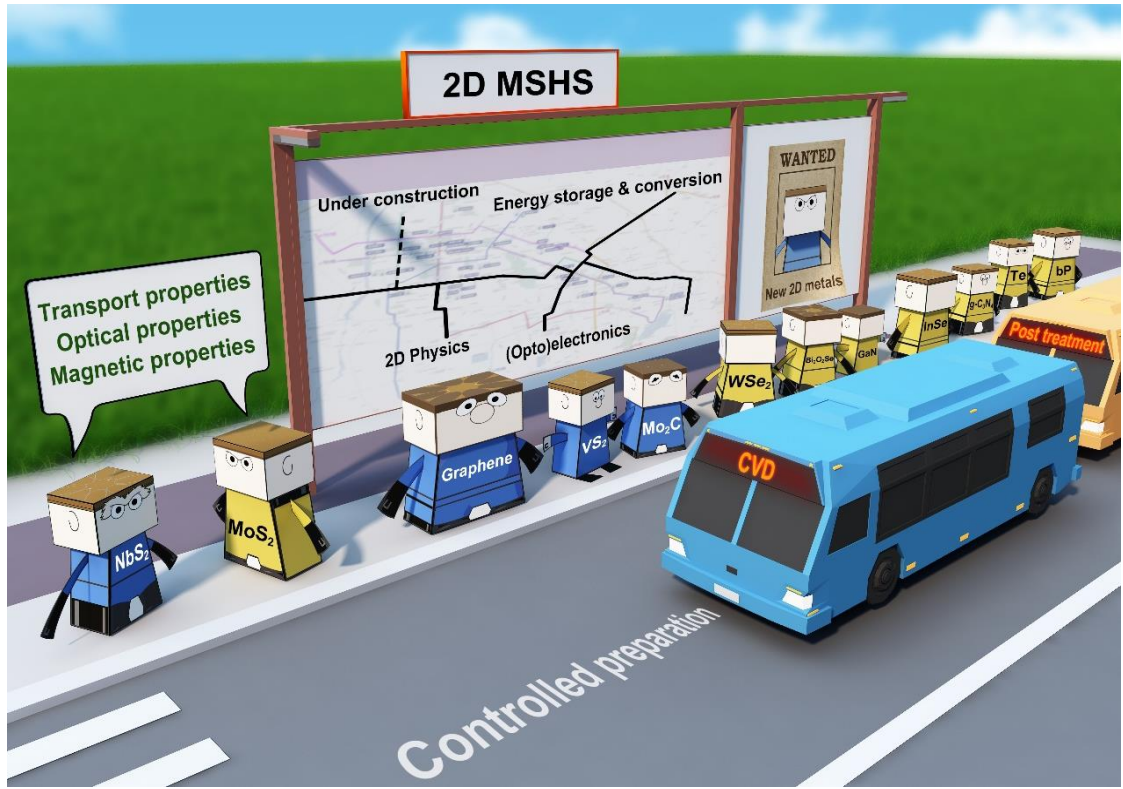

**Figure 8.** Outlook for 2D material-based MSHSs. For diverse future applications of MSHSs, controlled preparation methods (buses) are the foundation to realizing these goals. In addition, exploring more 2D metals (arrest order) and the interaction between 2D metals and semiconductors (dialog box) will be important topics in this emerging field.

**Acknowledgements**

We acknowledge support from the National Natural Science Foundation of China (51722206, 51920105002, 51991340, and 51991343), Guangdong Innovative and Entrepreneurial Research Team Program (2017ZT07C341), the Bureau of Industry and Information Technology of Shenzhen for the "2017 Graphene Manufacturing Innovation Center Project" (201901171523).

**Conflict of Interest**

The authors declare no conflict of interest.

**Reference**

[1] K. S. Novoselov, A. K. Geim, S. V. Morozov, D. Jiang, Y. Zhang, S. V. Dubonos, I. V. Grigorieva, A. A. Firsov, *Science* **2004**, *306*, 666.

[2] N. Mounet, M. Gibertini, P. Schwaller, D. Campi, A. Merkys, A. Marrazzo, T. Sohier, I. E. Castelli, A. Cepellotti, G. Pizzi, N. Marzari, *Nat Nanotechnol* **2018**, *13*, 246.

[3] C. Tan, X. Cao, X. J. Wu, Q. He, J. Yang, X. Zhang, J. Chen, W. Zhao, S. Han, G. H. Nam, M. Sindoro, H. Zhang, *Chem Rev* **2017**, *117*, 6225.

[4] K. S. Novoselov, D. Jiang, F. Schedin, T. J. Booth, V. V. Khotkevich, S. V. Morozov, A. K. Geim, *Proc Natl Acad Sci U S A* **2005**, *102*, 10451.

[5] Y. Liu, Y. Huang, X. Duan, *Nature* **2019**, *567*, 323.

[6] A. K. Geim, I. V. Grigorieva, *Nature* **2013**, *499*, 419.

[7] C. Tan, H. Zhang, *J Am Chem Soc* **2015**, *137*, 12162.

[8] M. Y. Li, C. H. Chen, Y. M. Shi, L. J. Li, *Mater Today* **2016**, *19*, 322.

[9] C. R. Dean, L. Wang, P. Maher, C. Forsythe, F. Ghahari, Y. Gao, J. Katoch, M. Ishigami, P. Moon, M. Koshino, T. Taniguchi, K. Watanabe, K. L. Shepard, J. Hone, P. Kim, *Nature* **2013**, *497*, 598.

[10] X. Hong, J. Kim, S. F. Shi, Y. Zhang, C. Jin, Y. Sun, S. Tongay, J. Wu, Y. Zhang, F. Wang, *Nat Nanotechnol* **2014**, *9*, 682.

[11] T. Georgiou, R. Jalil, B. D. Belle, L. Britnell, R. V. Gorbachev, S. V. Morozov, Y. J. Kim, A. Gholinia, S. J. Haigh, O. Makarovsky, L. Eaves, L. A. Ponomarenko, A. K. Geim, K. S. Novoselov, A. Mishchenko, *Nat Nanotechnol* **2013**, *8*, 100.

[12] T. Roy, M. Tosun, J. S. Kang, A. B. Sachid, S. B. Desai, M. Hettick, C. C. Hu, A. Javey, *ACS Nano* **2014**, *8*, 6259.

[13] W. Zhang, C. P. Chuu, J. K. Huang, C. H. Chen, M. L. Tsai, Y. H. Chang, C. T. Liang, Y. Z. Chen, Y. L. Chueh, J. H. He, M. Y. Chou, L. J. Li, *Sci Rep* **2014**, *4*, 3826.

[14] J. D. Mehew, S. Unal, E. Torres Alonso, G. F. Jones, S. Fadhil Ramadhan, M. F. Craciun, S. Russo, *Adv Mater* **2017**, *29*, 1700222.

[15] F. Withers, O. Del Pozo-Zamudio, A. Mishchenko, A. P. Rooney, A. Gholinia, K. Watanabe, T. Taniguchi, S. J. Haigh, A. K. Geim, A. I. Tartakovskii, K. S. Novoselov, *Nat Mater* **2015**, *14*, 301.

[16] C. H. Liu, G. Clark, T. Fryett, S. Wu, J. Zheng, F. Hatami, X. Xu, A. Majumdar, *Nano Lett* **2017**, *17*, 200.

[17] G. H. Han, D. L. Duong, D. H. Keum, S. J. Yun, Y. H. Lee, *Chem Rev* **2018**, *118*, 6297.

[18] B. Anasori, M. R. Lukatskaya, Y. Gogotsi, *Nat Rev Mater* **2017**, *2*, 16098.

[19] J. H. Kang, W. Liu, D. Sarkar, D. Jena, K. Banerjee, *Phys. Rev. X* **2014**, *4*, 031005.

[20] J. Du, C. Xia, W. Xiong, T. Wang, Y. Jia, J. Li, *Nanoscale* **2017**, *9*, 17585.

[21] A. Allain, J. Kang, K. Banerjee, A. Kis, *Nat Mater* **2015**, *14*, 1195.

[22] A. R. Kim, Y. Kim, J. Nam, H. S. Chung, D. J. Kim, J. D. Kwon, S. W. Park, J. Park, S. Y. Choi, B. H. Lee, J. H. Park, K. H. Lee, D. H. Kim, S. M. Choi, P. M. Ajayan, M. G. Hahm, B. Cho,

*Nano Lett* **2016**, *16*, 1890.

[23] H. J. Chuang, B. Chamlagain, M. Koehler, M. M. Perera, J. Yan, D. Mandrus, D. Tomanek, Z. Zhou, *Nano Lett* **2016**, *16*, 1896.

[24] D. Deng, K. S. Novoselov, Q. Fu, N. Zheng, Z. Tian, X. Bao, *Nat Nanotechnol* **2016**, *11*, 218.

[25] E. Pomerantseva, Y. Gogotsi, *Nat Energy* **2017**, *2*, 17089.

[26] J. Yuan, J. Wu, W. J. Hardy, P. Loya, M. Lou, Y. Yang, S. Najmaei, M. Jiang, F. Qin, K. Keyshar, H. Ji, W. Gao, J. Bao, J. Kono, D. Natelson, P. M. Ajayan, J. Lou, *Adv Mater* **2015**, *27*, 5605.

[27] Y. Saito, T. Nojima, Y. Iwasa, *Nat Rev Mater* **2016**, *2*, 16094.

[28] X. Xi, L. Zhao, Z. Wang, H. Berger, L. Forro, J. Shan, K. F. Mak, *Nat Nanotechnol* **2015**, *10*, 765.

[29] M. H. Guimaraes, H. Gao, Y. Han, K. Kang, S. Xie, C. J. Kim, D. A. Muller, D. C. Ralph, J. Park, *ACS Nano* **2016**, *10*, 6392.

[30] X. Ling, Y. Lin, Q. Ma, Z. Wang, Y. Song, L. Yu, S. Huang, W. Fang, X. Zhang, A. L. Hsu, Y. Bie, Y. H. Lee, Y. Zhu, L. Wu, J. Li, P. Jarillo-Herrero, M. Dresselhaus, T. Palacios, J. Kong, *Adv Mater* **2016**, *28*, 2322.

[31] C. H. Naylor, W. M. Parkin, Z. Gao, J. Berry, S. Zhou, Q. Zhang, J. B. McClimon, L. Z. Tan, C. E. Kehayias, M. Q. Zhao, R. S. Gona, R. W. Carpick, A. M. Rappe, D. J. Srolovitz, M. Drndic, A. T. C. Johnson, *ACS Nano* **2017**, *11*, 8619.

[32] J. Jeon, Y. Park, S. Choi, J. Lee, S. S. Lim, B. H. Lee, Y. J. Song, J. H. Cho, Y. H. Jang, S. Lee, *ACS Nano* **2018**, *12*, 338.

[33] S. D. Fan, Q. A. Vu, M. D. Tran, S. Adhikari, Y. H. Lee, *2d Mater* **2020**, *7*, 022005.

[34] Z. Li, R. Ye, R. Feng, Y. Kang, X. Zhu, J. M. Tour, Z. Fang, *Adv Mater* **2015**, *27*, 5235.

[35] W. Zhao, S. Wang, B. Liu, I. Verzhbitskiy, S. Li, F. Giustiniano, D. Kozawa, K. P. Loh, K. Matsuda, K. Okamoto, R. F. Oulton, G. Eda, *Adv Mater* **2016**, *28*, 2709.

[36] X. Huang, Z. Zeng, S. Bao, M. Wang, X. Qi, Z. Fan, H. Zhang, *Nat Commun* **2013**, *4*, 1444.

[37] S. B. Desai, S. R. Madhvapathy, A. B. Sachid, J. P. Llinas, Q. Wang, G. H. Ahn, G. Pitner, M. J. Kim, J. Bokor, C. Hu, H. P. Wong, A. Javey, *Science* **2016**, *354*, 99.

[38] L. Li, Y. Guo, Y. Sun, L. Yang, L. Qin, S. Guan, J. Wang, X. Qiu, H. Li, Y. Shang, Y. Fang, *Adv Mater* **2018**, *30*, 1706215.

[39] J. Zhang, K. Zhang, B. Xia, Y. Wei, D. Li, K. Zhang, Z. Zhang, Y. Wu, P. Liu, X. Duan, Y. Xu, W. Duan, S. Fan, K. Jiang, *Adv Mater* **2017**, *29*, 1702942.

[40] J. Zhang, Y. Wei, F. Yao, D. Li, H. Ma, P. Lei, H. Fang, X. Xiao, Z. Lu, J. Yang, J. Li, L. Jiao, W. Hu, K. Liu, K. Liu, P. Liu, Q. Li, W. Lu, S. Fan, K. Jiang, *Adv Mater* **2017**, *29*, 1604469.

[41] S. Tongay, M. Lemaitre, T. Schumann, K. Berke, B. R. Appleton, B. Gila, A. F. Hebard, *Appl Phys Lett* **2011**, *99*, 102102.

[42] L. Ci, L. Song, C. Jin, D. Jariwala, D. Wu, Y. Li, A. Srivastava, Z. F. Wang, K. Storr, L. Balicas, F. Liu, P. M. Ajayan, *Nat Mater* **2010**, *9*, 430.

[43] Z. Liu, L. Ma, G. Shi, W. Zhou, Y. Gong, S. Lei, X. Yang, J. Zhang, J. Yu, K. P. Hackenberg, A. Babakhani, J. C. Idrobo, R. Vajtai, J. Lou, P. M. Ajayan, *Nat Nanotechnol* **2013**, *8*, 119.

[44] M. P. Levendorf, C. J. Kim, L. Brown, P. Y. Huang, R. W. Havener, D. A. Muller, J. Park, *Nature* **2012**, *488*, 627.

[45] Y. Gong, J. Lin, X. Wang, G. Shi, S. Lei, Z. Lin, X. Zou, G. Ye, R. Vajtai, B. I. Yakobson, H. Terrones, M. Terrones, B. K. Tay, J. Lou, S. T. Pantelides, Z. Liu, W. Zhou, P. M. Ajayan, *Nat Mater* **2014**, *13*, 1135.

[46] K. Chen, X. Wan, W. Xie, J. Wen, Z. Kang, X. Zeng, H. Chen, J. Xu, *Adv Mater* **2015**, *27*, 6431.

[47] C. Huang, S. Wu, A. M. Sanchez, J. J. Peters, R. Beanland, J. S. Ross, P. Rivera, W. Yao, D. H. Cobden, X. Xu, *Nat Mater* **2014**, *13*, 1096.

[48] X. Q. Zhang, C. H. Lin, Y. W. Tseng, K. H. Huang, Y. H. Lee, *Nano Lett* **2015**, *15*, 410.

[49] X. Chen, Y. Qiu, H. Yang, G. Liu, W. Zheng, W. Feng, W. Cao, W. Hu, P. Hu, *ACS Appl Mater Interfaces* **2017**, *9*, 1684.

[50] M. Y. Li, Y. Shi, C. C. Cheng, L. S. Lu, Y. C. Lin, H. L. Tang, M. L. Tsai, C. W. Chu, K. H. Wei, J. H. He, W. H. Chang, K. Suenaga, L. J. Li, *Science* **2015**, *349*, 524.

[51] M. Y. Li, J. Pu, J. K. Huang, Y. Miyauchi, K. Matsuda, T. Takenobu, L. J. Li, *Adv Funct Mater* **2018**, *28*, 1706860.

[52] D. Liu, J. Hong, X. Wang, X. Li, Q. Feng, C. Tan, T. Zhai, F. Ding, H. Peng, H. Xu, *Adv Funct Mater* **2018**, *28*, 1804696.

[53] D. Liu, J. Hong, X. Li, X. Zhou, B. Jin, Q. Cui, J. Chen, Q. Feng, C. Xu, T. Zhai, K. Suenaga, H. Xu, *Adv Funct Mater* **2020**, *30*, 1910169.

[54] G. Kim, H. S. Shin, *Nanoscale* **2020**, *12*, 5286.

[55] M. Zhao, Y. Ye, Y. Han, Y. Xia, H. Zhu, S. Wang, Y. Wang, D. A. Muller, X. Zhang, *Nat Nanotechnol* **2016**, *11*, 954.

[56] H. L. Tang, M. H. Chiu, C. C. Tseng, S. H. Yang, K. J. Hou, S. Y. Wei, J. K. Huang, Y. F. Lin, C. H. Lien, L. J. Li, *ACS Nano* **2017**, *11*, 12817.

[57] C. Zheng, Q. Zhang, B. Weber, H. Ilatikhameneh, F. Chen, H. Sahasrabudhe, R. Rahman, S. Li, Z. Chen, J. Hellerstedt, Y. Zhang, W. H. Duan, Q. Bao, M. S. Fuhrer, *ACS Nano* **2017**, *11*, 2785.

[58] W. Hong, G. W. Shim, S. Y. Yang, D. Y. Jung, S. Y. Choi, *Adv Funct Mater* **2018**, *29*, 1807550.

[59] S. S. Li, S. F. Wang, D. M. Tang, W. J. Zhao, H. L. Xu, L. Q. Chu, Y. Bando, D. Golberg, G. Eda, *Appl Mater Today* **2015**, *1*, 60.

[60] J. Zhou, J. Lin, X. Huang, Y. Zhou, Y. Chen, J. Xia, H. Wang, Y. Xie, H. Yu, J. Lei, D. Wu, F. Liu, Q. Fu, Q. Zeng, C. H. Hsu, C. Yang, L. Lu, T. Yu, Z. Shen, H. Lin, B. I. Yakobson, Q. Liu, K. Suenaga, G. Liu, Z. Liu, *Nature* **2018**, *556*, 355.

[61] Z. Wang, Y. Xie, H. Wang, R. Wu, T. Nan, Y. Zhan, J. Sun, T. Jiang, Y. Zhao, Y. Lei, M. Yang, W. Wang, Q. Zhu, X. Ma, Y. Hao, *Nanotechnology* **2017**, *28*, 325602.

[62] J. Zhou, B. Tang, J. Lin, D. Lv, J. Shi, L. Sun, Q. Zeng, L. Niu, F. Liu, X. Wang, X. Liu, K. Suenaga, C. Jin, Z. Liu, *Adv Funct Mater* **2018**, *28*, 1801568.

[63] A. Apte, A. Krishnamoorthy, J. A. Hachtel, S. Susarla, J. Yoon, L. M. Sassi, P. Bharadwaj, J. M. Tour, J. C. Idrobo, R. K. Kalia, A. Nakano, P. Vashishta, C. S. Tiwary, P. M. Ajayan, *Nano Lett* **2019**, *19*, 6338.

[64] Y. Zhang, L. Yin, J. Chu, T. A. Shifa, J. Xia, F. Wang, Y. Wen, X. Zhan, Z. Wang, J. He, *Adv Mater* **2018**, *30*, 1803665.

[65] X. Gong, X. Zhao, M. E. Pam, H. Yao, Z. Li, D. Geng, S. J. Pennycook, Y. Shi, H. Y. Yang, *Nanoscale* **2019**, *11*, 4183.

[66] W. S. Leong, Q. Ji, N. Mao, Y. Han, H. Wang, A. J. Goodman, A. Vignon, C. Su, Y. Guo, P. C. Shen, Z. Gao, D. A. Muller, W. A. Tisdale, J. Kong, *J Am Chem Soc* **2018**, *140*, 12354.

[67] Z. Zhang, P. Chen, X. Duan, K. Zang, J. Luo, X. Duan, *Science* **2017**, *357*, 788.

[68] P. K. Sahoo, S. Memaran, Y. Xin, L. Balicas, H. R. Gutierrez, *Nature* **2018**, *553*, 63.

[69] J. H. Sung, H. Heo, S. Si, Y. H. Kim, H. R. Noh, K. Song, J. Kim, C. S. Lee, S. Y. Seo, D. H. Kim, H. K. Kim, H. W. Yeom, T. H. Kim, S. Y. Choi, J. S. Kim, M. H. Jo, *Nat Nanotechnol* **2017**, *12*, 1064.

[70] Y. Yoo, Z. P. DeGregorio, Y. Su, S. J. Koester, J. E. Johns, *Adv Mater* **2017**, *29*, 1605461.

[71] R. Ma, H. Zhang, Y. Yoo, Z. P. Degregorio, L. Jin, P. Golani, J. Ghasemi Azadani, T. Low, J. E. Johns, L. A. Bendersky, A. V. Davydov, S. J. Koester, *ACS Nano* **2019**, *13*, 8035.

[72] X. Xu, S. Chen, S. Liu, X. Cheng, W. Xu, P. Li, Y. Wan, S. Yang, W. Gong, K. Yuan, P. Gao, Y. Ye, L. Dai, *J Am Chem Soc* **2019**, *141*, 2128.

[73] X. Xu, S. Liu, B. Han, Y. Han, K. Yuan, W. Xu, X. Yao, P. Li, S. Yang, W. Gong, D. A. Muller, P. Gao, Y. Ye, L. Dai, *Nano Lett* **2019**, *19*, 6845.

[74] D. H. Keum, S. Cho, J. H. Kim, D. H. Choe, H. J. Sung, M. Kan, H. Kang, J. Y. Hwang, S. W. Kim, H. Yang, K. J. Chang, Y. H. Lee, *Nat Phys* **2015**, *11*, 482.

[75] Q. Zhang, X. F. Wang, S. H. Shen, Q. Lu, X. Z. Liu, H. Y. Li, J. Y. Zheng, C. P. Yu, X. Y. Zhong, L. Gu, T. L. Ren, L. Y. Jiao, *Nat Electron* **2019**, *2*, 164.

[76] C. Ataca, H. Sahin, S. Ciraci, *J Phys Chem C* **2012**, *116*, 8983.

[77] Q. Ji, C. Li, J. Wang, J. Niu, Y. Gong, Z. Zhang, Q. Fang, Y. Zhang, J. Shi, L. Liao, X. Wu, L. Gu, Z. Liu, Y. Zhang, *Nano Lett* **2017**, *17*, 4908.

[78] J. Su, M. Wang, Y. Li, F. Wang, Q. Chen, P. Luo, J. Han, S. Wang, H. Li, T. Zhai, *Adv Funct Mater* **2020**, *30*, 2000240.

[79] Y. Huan, J. Shi, X. Zou, Y. Gong, Z. Zhang, M. Li, L. Zhao, R. Xu, S. Jiang, X. Zhou, M. Hong, C. Xie, H. Li, X. Lang, Q. Zhang, L. Gu, X. Yan, Y. Zhang, *Adv Mater* **2018**, *30*, 1705916.

[80] J. Shi, X. Chen, L. Zhao, Y. Gong, M. Hong, Y. Huan, Z. Zhang, P. Yang, Y. Li, Q. Zhang, Q. Zhang, L. Gu, H. Chen, J. Wang, S. Deng, N. Xu, Y. Zhang, *Adv Mater* **2018**, *30*, 1804616.

[81] A. N. Enyashin, L. Yadgarov, L. Houben, I. Popov, M. Weidenbach, R. Tenne, M. Bar-Sadan, G. Seifert, *J Phys Chem C* **2011**, *115*, 24586.

[82] Y. H. Lee, X. Q. Zhang, W. Zhang, M. T. Chang, C. T. Lin, K. D. Chang, Y. C. Yu, J. T. Wang, C. S. Chang, L. J. Li, T. W. Lin, *Adv Mater* **2012**, *24*, 2320.

[83] X. Ling, Y. H. Lee, Y. X. Lin, W. J. Fang, L. L. Yu, M. S. Dresselhaus, J. Kong, *Nano Lett* **2014**, *14*, 464.

[84] J. Chen, W. Tang, B. Tian, B. Liu, X. Zhao, Y. Liu, T. Ren, W. Liu, D. Geng, H. Y. Jeong, H. S. Shin, W. Zhou, K. P. Loh, *Adv Sci* **2016**, *3*, 1500033.

[85] Y. Yu, G. H. Nam, Q. He, X. J. Wu, K. Zhang, Z. Yang, J. Chen, Q. Ma, M. Zhao, Z. Liu, F. R. Ran, X. Wang, H. Li, X. Huang, B. Li, Q. Xiong, Q. Zhang, Z. Liu, L. Gu, Y. Du, W. Huang, H. Zhang, *Nat Chem* **2018**, *10*, 638.

[86] L. Liu, J. Wu, L. Wu, M. Ye, X. Liu, Q. Wang, S. Hou, P. Lu, L. Sun, J. Zheng, L. Xing, L. Gu, X. Jiang, L. Xie, L. Jiao, *Nat Mater* **2018**, *17*, 1108.

[87] Y. C. Lin, C. H. Yeh, H. C. Lin, M. D. Siao, Z. Liu, H. Nakajima, T. Okazaki, M. Y. Chou, K. Suenaga, P. W. Chiu, *ACS Nano* **2018**, *12*, 12080.

[88] M. Kan, J. Y. Wang, X. W. Li, S. H. Zhang, Y. W. Li, Y. Kawazoe, Q. Sun, P. Jena, *J Phys Chem C* **2014**, *118*, 1515.

[89] M. A. Py, R. R. Haering, *Can J Phy* **1983**, *61*, 76.

[90] G. Gao, Y. Jiao, F. Ma, Y. Jiao, E. Waclawik, A. Du, *J Phys Chem C* **2015**, *119*, 13124.

[91] S. S. Chou, N. Sai, P. Lu, E. N. Coker, S. Liu, K. Artyushkova, T. S. Luk, B. Kaehr, C. J. Brinker, *Nat Commun* **2015**, *6*, 8311.

[92]  K. Leng, Z. Chen, X. Zhao, W. Tang, B. Tian, C. T. Nai, W. Zhou, K. P. Loh, *ACS Nano* **2016**, *10*, 9208.

[93]  J. Wu, J. Peng, Y. Zhou, Y. Lin, X. Wen, J. Wu, Y. Zhao, Y. Guo, C. Wu, Y. Xie, *J Am Chem Soc* **2019**, *141*, 592.

[94]  R. Kappera, D. Voiry, S. E. Yalcin, B. Branch, G. Gupta, A. D. Mohite, M. Chhowalla, *Nat Mater* **2014**, *13*, 1128.

[95]  Y. Ma, B. Liu, A. Zhang, L. Chen, M. Fathi, C. Shen, A. N. Abbas, M. Ge, M. Mecklenburg, C. Zhou, *ACS Nano* **2015**, *9*, 7383.

[96]  L. Sun, X. Yan, J. Zheng, H. Yu, Z. Lu, S. P. Gao, L. Liu, X. Pan, D. Wang, Z. Wang, P. Wang, L. Jiao, *Nano Lett* **2018**, *18*, 3435.

[97]  Y. Gong, H. Yuan, C. L. Wu, P. Tang, S. Z. Yang, A. Yang, G. Li, B. Liu, J. van de Groep, M. L. Brongersma, M. F. Chisholm, S. C. Zhang, W. Zhou, Y. Cui, *Nat Nanotechnol* **2018**, *13*, 294.

[98]  Y. C. Lin, D. O. Dumcenco, Y. S. Huang, K. Suenaga, *Nat Nanotechnol* **2014**, *9*, 391.

[99]  J. Lin, O. Cretu, W. Zhou, K. Suenaga, D. Prasai, K. I. Bolotin, N. T. Cuong, M. Otani, S. Okada, A. R. Lupini, J. C. Idrobo, D. Caudel, A. Burger, N. J. Ghimire, J. Yan, D. G. Mandrus, S. J. Pennycook, S. T. Pantelides, *Nat Nanotechnol* **2014**, *9*, 436.

[100]  L. Tao, X. Duan, C. Wang, X. Duan, S. Wang, *Chem Commun* **2015**, *51*, 7470.

[101]  H. Li, Y. Tan, P. Liu, C. Guo, M. Luo, J. Han, T. Lin, F. Huang, M. Chen, *Adv Mater* **2016**, *28*, 8945.

[102]  A. Castellanos-Gomez, M. Barkelid, A. M. Goossens, V. E. Calado, H. S. van der Zant, G. A. Steele, *Nano Lett* **2012**, *12*, 3187.

[103]  Y. Liu, H. Nan, X. Wu, W. Pan, W. Wang, J. Bai, W. Zhao, L. Sun, X. Wang, Z. Ni, *ACS Nano* **2013**, *7*, 4202.

[104]  J. Zhu, Z. Wang, H. Yu, N. Li, J. Zhang, J. Meng, M. Liao, J. Zhao, X. Lu, L. Du, R. Yang, D. Shi, Y. Jiang, G. Zhang, *J Am Chem Soc* **2017**, *139*, 10216.

[105]  J. H. Kim, S. J. Yun, H. S. Lee, J. Zhao, H. Bouzid, Y. H. Lee, *Sci Rep* **2018**, *8*, 10284.

[106]  O. Baranov, I. Levchenko, J. M. Bell, J. W. M. Lim, S. Huang, L. Xu, B. Wang, D. U. B. Aussems, S. Xu, K. Bazaka, *Mater Horiz* **2018**, *5*, 765.

[107]  X. Lin, J. C. Lu, Y. Shao, Y. Y. Zhang, X. Wu, J. B. Pan, L. Gao, S. Y. Zhu, K. Qian, Y. F. Zhang, D. L. Bao, L. F. Li, Y. Q. Wang, Z. L. Liu, J. T. Sun, T. Lei, C. Liu, J. O. Wang, K. Ibrahim, D. N. Leonard, W. Zhou, H. M. Guo, Y. L. Wang, S. X. Du, S. T. Pantelides, H. J. Gao, *Nat Mater* **2017**, *16*, 717.

[108]  S. Susarla, J. A. Hachtel, X. Yang, A. Kutana, A. Apte, Z. Jin, R. Vajtai, J. C. Idrobo, J. Lou, B. I. Yakobson, C. S. Tiwary, P. M. Ajayan, *Adv Mater* **2018**, *30*, 1804218.

[109]  X. Yin, C. S. Tang, D. Wu, W. Kong, C. Li, Q. Wang, L. Cao, M. Yang, Y. H. Chang, D. Qi, F. Ouyang, S. J. Pennycook, Y. P. Feng, M. B. H. Breese, S. J. Wang, W. Zhang, A. Rusydi, A. T. S. Wee, *Adv Sci* **2019**, *6*, 1802093.

[110]  H. Zhu, Q. Wang, C. Zhang, R. Addou, K. Cho, R. M. Wallace, M. J. Kim, *Adv Mater* **2017**, *29*, 1606264.

[111]  H. Wang, Y. Wu, X. Yuan, G. Zeng, J. Zhou, X. Wang, J. W. Chew, *Adv Mater* **2018**, *30*, 1704561.

[112]  M. Naguib, M. Kurtoglu, V. Presser, J. Lu, J. Niu, M. Heon, L. Hultman, Y. Gogotsi, M. W. Barsoum, *Adv Mater* **2011**, *23*, 4248.

[113]  Y. Li, H. Shao, Z. Lin, J. Lu, L. Liu, B. Duployer, P. O. A. Persson, P. Eklund, L. Hultman, M. Li, K. Chen, X. H. Zha, S. Du, P. Rozier, Z. Chai, E. Raymundo-Pinero, P. L. Taberna, P. Simon, Q. Huang, *Nat Mater* **2020**, *19*, 894.

[114]  Y. Luo, L. Tang, U. Khan, Q. Yu, H. M. Cheng, X. Zou, B. Liu, *Nat Commun* **2019**, *10*, 269.

[115]  Z. Du, S. Yang, S. Li, J. Lou, S. Zhang, S. Wang, B. Li, Y. Gong, L. Song, X. Zou, P. M. Ajayan, *Nature* **2020**, *577*, 492.

[116]  S. Choi, Y. J. Kim, J. Jeon, B. H. Lee, J. H. Cho, S. Lee, *ACS Appl Mater Interfaces* **2019**, *11*, 47190.

[117]  J. Cao, T. Li, H. Gao, Y. Lin, X. Wang, H. Wang, T. Palacios, X. Ling, *Sci Adv* **2020**, *6*, eaax8784.

[118]  Y. Liu, N. O. Weiss, X. D. Duan, H. C. Cheng, Y. Huang, X. F. Duan, *Nat Rev Mater* **2016**, *1*, 16042.

[119]  Y. Huang, E. Sutter, N. N. Shi, J. Zheng, T. Yang, D. Englund, H. J. Gao, P. Sutter, *ACS Nano* **2015**, *9*, 10612.

[120]  S. B. Desai, S. R. Madhvapathy, M. Amani, D. Kiriya, M. Hettick, M. Tosun, Y. Zhou, M. Dubey, J. W. Ager, 3rd, D. Chrzan, A. Javey, *Adv Mater* **2016**, *28*, 4053.

[121]  F. Liu, W. Wu, Y. Bai, S. H. Chae, Q. Li, J. Wang, J. Hone, X. Y. Zhu, *Science* **2020**, *367*, 903.

[122]  J. Shim, S. H. Bae, W. Kong, D. Lee, K. Qiao, D. Nezich, Y. J. Park, R. Zhao, S. Sundaram, X. Li, H. Yeon, C. Choi, H. Kum, R. Yue, G. Zhou, Y. Ou, K. Lee, J. Moodera, X. Zhao, J. H. Ahn, C. Hinkle, A. Ougazzaden, J. Kim, *Science* **2018**, *362*, 665.

[123]  K. Kang, K. H. Lee, Y. Han, H. Gao, S. Xie, D. A. Muller, J. Park, *Nature* **2017**, *550*, 229.

[124]  M. Liao, Z. Wei, L. Du, Q. Wang, J. Tang, H. Yu, F. Wu, J. Zhao, X. Xu, B. Han, K. Liu, P. Gao, T. Polcar, Z. Sun, D. Shi, R. Yang, G. Zhang, *Nat Commun* **2020**, *11*, 2153.

[125]  H. Chen, X. L. Zhang, Y. Y. Zhang, D. Wang, D. L. Bao, Y. Que, W. Xiao, S. Du, M. Ouyang, S. T. Pantelides, H. J. Gao, *Science* **2019**, *365*, 1036.

[126]  Y. Wakafuji, R. Moriya, S. Masubuchi, K. Watanabe, T. Taniguchi, T. Machida, *Nano Lett* **2020**, *20*, 2486.

[127]  A. Koma, *Thin Solid Films* **1992**, *216*, 72.

[128]  Y. Zhang, T. R. Chang, B. Zhou, Y. T. Cui, H. Yan, Z. Liu, F. Schmitt, J. Lee, R. Moore, Y. Chen, H. Lin, H. T. Jeng, S. K. Mo, Z. Hussain, A. Bansil, Z. X. Shen, *Nat Nanotechnol* **2014**, *9*, 111.

[129]  A. Azizi, S. Eichfeld, G. Geschwind, K. Zhang, B. Jiang, D. Mukherjee, L. Hossain, A. F. Piasecki, B. Kabius, J. A. Robinson, N. Alem, *ACS Nano* **2015**, *9*, 4882.

[130]  Y. Shi, W. Zhou, A. Y. Lu, W. Fang, Y. H. Lee, A. L. Hsu, S. M. Kim, K. K. Kim, H. Y. Yang, L. J. Li, J. C. Idrobo, J. Kong, *Nano Lett* **2012**, *12*, 2784.

[131]  J. Shi, M. Liu, J. Wen, X. Ren, X. Zhou, Q. Ji, D. Ma, Y. Zhang, C. Jin, H. Chen, S. Deng, N. Xu, Z. Liu, Y. Zhang, *Adv Mater* **2015**, *27*, 7086.

[132]  Y. C. Lin, R. K. Ghosh, R. Addou, N. Lu, S. M. Eichfeld, H. Zhu, M. Y. Li, X. Peng, M. J. Kim, L. J. Li, R. M. Wallace, S. Datta, J. A. Robinson, *Nat Commun* **2015**, *6*, 7311.

[133]  Q. D. Fu, X. W. Wang, J. D. Zhou, J. Xia, Q. S. Zeng, D. H. Lv, C. Zhu, X. L. Wang, Y. Shen, X. M. Li, Y. N. Hua, F. C. Liu, Z. X. Shen, C. H. Jin, Z. Liu, *Chem Mater* **2018**, *30*, 4001.

[134]  C. S. Lee, S. J. Oh, H. Heo, S. Y. Seo, J. Kim, Y. H. Kim, D. Kim, O. F. Ngome Okello, H. Shin, J. H. Sung, S. Y. Choi, J. S. Kim, J. K. Kim, M. H. Jo, *Nano Lett* **2019**, *19*, 1814.

[135]  R. Wu, Q. Tao, W. Dang, Y. Liu, B. Li, J. Li, B. Zhao, Z. Zhang, H. Ma, G. Sun, X. Duan, X. Duan, *Adv Funct Mater* **2019**, *29*, 1806611.

[136]  Z. Zhang, Y. Gong, X. Zou, P. Liu, P. Yang, J. Shi, L. Zhao, Q. Zhang, L. Gu, Y. Zhang, *ACS*

*Nano* **2019**, *13*, 885.

[137] Z. Wang, R. Luo, I. Johnson, H. Kashani, M. Chen, *ACS Nano* **2020**, *14*, 899.

[138] Y. Han, M. Y. Li, G. S. Jung, M. A. Marsalis, Z. Qin, M. J. Buehler, L. J. Li, D. A. Muller, *Nat Mater* **2018**, *17*, 129.

[139] W. Zhou, Y. Y. Zhang, J. Chen, D. Li, J. Zhou, Z. Liu, M. F. Chisholm, S. T. Pantelides, K. P. Loh, *Sci Adv* **2018**, *4*, eaap9096.

[140] C. Zhu, M. Yu, J. Zhou, Y. He, Q. Zeng, Y. Deng, S. Guo, M. Xu, J. Shi, W. Zhou, L. Sun, L. Wang, Z. Hu, Z. Zhang, W. Guo, Z. Liu, *Nat Commun* **2020**, *11*, 772.

[141] J. Li, X. Yang, Y. Liu, B. Huang, R. Wu, Z. Zhang, B. Zhao, H. Ma, W. Dang, Z. Wei, K. Wang, Z. Lin, X. Yan, M. Sun, B. Li, X. Pan, J. Luo, G. Zhang, Y. Liu, Y. Huang, X. Duan, X. Duan, *Nature* **2020**, *579*, 368.

[142] R. Xiang, T. Inoue, Y. Zheng, A. Kumamoto, Y. Qian, Y. Sato, M. Liu, D. Tang, D. Gokhale, J. Guo, K. Hisama, S. Yotsumoto, T. Ogamoto, H. Arai, Y. Kobayashi, H. Zhang, B. Hou, A. Anisimov, M. Maruyama, Y. Miyata, S. Okada, S. Chiashi, Y. Li, J. Kong, E. I. Kauppinen, Y. Ikuhara, K. Suenaga, S. Maruyama, *Science* **2020**, *367*, 537.

[143] V. T. Nguyen, W. Yim, S. J. Park, B. H. Son, Y. C. Kim, T. T. Cao, Y. Sim, Y. J. Moon, V. C. Nguyen, M. J. Seong, S. K. Kim, Y. H. Ahn, S. Lee, J. Y. Park, *Adv Funct Mater* **2018**, *28*, 1802572.

[144] R. Wang, T. Wang, T. Hong, Y. Q. Xu, *Nanotechnology* **2018**, *29*, 345205.

[145] Y. Gao, Z. Liu, D. M. Sun, L. Huang, L. P. Ma, L. C. Yin, T. Ma, Z. Zhang, X. L. Ma, L. M. Peng, H. M. Cheng, W. Ren, *Nat Commun* **2015**, *6*, 8569.

[146] Y. Zhang, J. Shi, G. Han, M. Li, Q. Ji, D. Ma, Y. Zhang, C. Li, X. Lang, Y. Zhang, Z. Liu, *Nano Research* **2015**, *8*, 2881.

[147] Y. Gao, Y. L. Hong, L. C. Yin, Z. Wu, Z. Yang, M. L. Chen, Z. Liu, T. Ma, D. M. Sun, Z. Ni, X. L. Ma, H. M. Cheng, W. Ren, *Adv Mater* **2017**, *29*, 1700990.

[148] R. Luo, W. W. Xu, Y. Zhang, Z. Wang, X. Wang, Y. Gao, P. Liu, M. Chen, *Nat Commun* **2020**, *11*, 1011.

[149] H. E. Lim, T. Irisawa, N. Okada, M. Okada, T. Endo, Y. Nakanishi, Y. Maniwa, Y. Miyata, *Nanoscale* **2019**, *11*, 19700.

[150] X. Cai, Y. Luo, B. Liu, H. M. Cheng, *Chem Soc Rev* **2018**, *47*, 6224.

[151] C. Zhang, J. Y. Tan, Y. K. Pan, X. K. Cai, X. L. Zou, H. M. Cheng, B. L. Liu, *Natl Sci Rev* **2020**, *7*, 324.

[152] Z. Lin, Y. Liu, U. Halim, M. Ding, Y. Liu, Y. Wang, C. Jia, P. Chen, X. Duan, C. Wang, F. Song, M. Li, C. Wan, Y. Huang, X. Duan, *Nature* **2018**, *562*, 254.

[153] W. Yu, J. Li, T. S. Herng, Z. S. Wang, X. X. Zhao, X. Chi, W. Fu, I. Abdelwahab, J. Zhou, J. D. Dan, Z. X. Chen, Z. Chen, Z. J. Li, J. Lu, S. J. Pennycook, Y. P. Feng, J. Ding, K. P. Loh, *Adv Mater* **2019**, *31*, 1903779.

[154] M. C. Watts, L. Picco, F. S. Russell-Pavier, P. L. Cullen, T. S. Miller, S. P. Bartus, O. D. Payton, N. T. Skipper, V. Tileli, C. A. Howard, *Nature* **2019**, *568*, 216.

[155] G. Hu, J. Kang, L. W. T. Ng, X. Zhu, R. C. T. Howe, C. G. Jones, M. C. Hersam, T. Hasan, *Chem Soc Rev* **2018**, *47*, 3265.

[156] D. McManus, S. Vranic, F. Withers, V. Sanchez-Romaguera, M. Macucci, H. Yang, R. Sorrentino, K. Parvez, S. K. Son, G. Iannaccone, K. Kostarelos, G. Fiori, C. Casiraghi, *Nat Nanotechnol* **2017**, *12*, 343.

[157] T. Lv, Y. Yao, N. Li, T. Chen, *Angew Chem Int Ed* **2016**, *55*, 9191.

[158] L. Jiang, B. Lin, X. Li, X. Song, H. Xia, L. Li, H. Zeng, *ACS Appl Mater Interfaces* **2016**, *8*, 2680.

[159] M. R. Gao, J. X. Liang, Y. R. Zheng, Y. F. Xu, J. Jiang, Q. Gao, J. Li, S. H. Yu, *Nat Commun* **2015**, *6*, 5982.

[160] X. Wang, Z. Wang, J. Zhang, X. Wang, Z. Zhang, J. Wang, Z. Zhu, Z. Li, Y. Liu, X. Hu, J. Qiu, G. Hu, B. Chen, N. Wang, Q. He, J. Chen, J. Yan, W. Zhang, T. Hasan, S. Li, H. Li, H. Zhang, Q. Wang, X. Huang, W. Huang, *Nat Commun* **2018**, *9*, 3611.

[161] X. Zhou, Y. Liu, H. Ju, B. Pan, J. Zhu, T. Ding, C. Wang, Q. Yang, *Chem Mater* **2016**, *28*, 1838.

[162] Y. Sun, Y. Wang, J. Y. C. Chen, K. Fujisawa, C. F. Holder, J. T. Miller, V. H. Crespi, M. Terrones, R. E. Schaak, *Nat Chem* **2020**, *12*, 284.

[163] S. Cho, S. Kim, J. H. Kim, J. Zhao, J. Seok, D. H. Keum, J. Baik, D. H. Choe, K. J. Chang, K. Suenaga, S. W. Kim, Y. H. Lee, H. Yang, *Science* **2015**, *349*, 625.

[164] A. D. Oyedele, S. Yang, T. Feng, A. V. Haglund, Y. Gu, A. A. Puretzky, D. Briggs, C. M. Rouleau, M. F. Chisholm, R. R. Unocic, D. Mandrus, H. M. Meyer, 3rd, S. T. Pantelides, D. B. Geohegan, K. Xiao, *J Am Chem Soc* **2019**, *141*, 8928.

[165] V. Shautsova, S. Sinha, L. Hou, Q. Zhang, M. Tweedie, Y. Lu, Y. Sheng, B. F. Porter, H. Bhaskaran, J. H. Warner, *ACS Nano* **2019**, *13*, 14162.

[166] D. Akinwande, C. Huyghebaert, C.-H. Wang, M. I. Serna, S. Goossens, L.-J. Li, H. S. P. Wong, F. H. L. Koppens, *Nature* **2019**, *573*, 507.

[167] K. Kaasbjerg, K. S. Thygesen, K. W. Jacobsen, *Phys. Rev. B* **2012**, *85*, 115317.

[168] G. S. Kim, S. H. Kim, J. Park, K. H. Han, J. Kim, H. Y. Yu, *ACS Nano* **2018**, *12*, 6292.

[169] X. Wang, H. Feng, Y. Wu, L. Jiao, *J Am Chem Soc* **2013**, *135*, 5304.

[170] D. S. Schulman, A. J. Arnold, S. Das, *Chem Soc Rev* **2018**, *47*, 3037.

[171] K. L. Liu, P. Luo, W. Han, S. J. Yang, S. S. Zhou, H. Q. Li, T. Y. Zhai, *Sci Bull* **2019**, *64*, 1426.

[172] W. Schottky, *Zeitschrift für Physik A Hadrons and nuclei* **1939**, *113*, 367.

[173] N. F. Mott, *Proc R Soc Lon Ser-A* **1939**, *171*, 0027.

[174] H. Hasegawa, T. Sawada, *Thin Solid Films* **1983**, *103*, 119.

[175] R. T. Tung, *Phys Rev Lett* **2000**, *84*, 6078.

[176] C. Kim, I. Moon, D. Lee, M. S. Choi, F. Ahmed, S. Nam, Y. Cho, H. J. Shin, S. Park, W. J. Yoo, *ACS Nano* **2017**, *11*, 1588.

[177] Y. Liu, J. Guo, E. Zhu, L. Liao, S. J. Lee, M. Ding, I. Shakir, V. Gambin, Y. Huang, X. Duan, *Nature* **2018**, *557*, 696.

[178] Y. Liu, P. Stradins, S. H. Wei, *Sci Adv* **2016**, *2*, e1600069.

[179] S. Das, R. Gulotty, A. V. Sumant, A. Roelofs, *Nano Lett* **2014**, *14*, 2861.

[180] Y. Wang, J. C. Kim, R. J. Wu, J. Martinez, X. Song, J. Yang, F. Zhao, A. Mkhoyan, H. Y. Jeong, M. Chhowalla, *Nature* **2019**, *568*, 70.

[181] Z. Q. Fan, X. W. Jiang, J. Chen, J. W. Luo, *ACS Appl Mater Interfaces* **2018**, *10*, 19271.

[182] S. Liu, J. Li, B. Shi, X. Zhang, Y. Pan, M. Ye, R. Quhe, Y. Wang, H. Zhang, J. Yan, L. Xu, Y. Guo, F. Pan, J. Lu, *J Mater Chem C* **2018**, *6*, 5651.

[183] W. J. Yu, Y. Liu, H. Zhou, A. Yin, Z. Li, Y. Huang, X. Duan, *Nat Nanotechnol* **2013**, *8*, 952.

[184] A. Li, Q. Chen, P. Wang, Y. Gan, T. Qi, P. Wang, F. Tang, J. Z. Wu, R. Chen, L. Zhang, Y. Gong, *Adv Mater* **2019**, *31*, 1805656.

[185] R. R. Nair, P. Blake, A. N. Grigorenko, K. S. Novoselov, T. J. Booth, T. Stauber, N. M. Peres, A. K. Geim, *Science* **2008**, *320*, 1308.

[186] C. R. Zhu, D. Gao, J. Ding, D. Chao, J. Wang, *Chem Soc Rev* **2018**, *47*, 4332.

[187] Q. C. Wang, Y. P. Lei, Y. C. Wang, Y. Liu, C. Y. Song, J. Zeng, Y. H. Song, X. D. Duan, D. S. Wang, Y. D. Li, *Energ Environ Sci* **2020**, *13*, 1593.

[188] X. Wang, Q. Weng, Y. Yang, Y. Bando, D. Golberg, *Chem Soc Rev* **2016**, *45*, 4042.

[189] S. Chandrasekaran, L. Yao, L. Deng, C. Bowen, Y. Zhang, S. Chen, Z. Lin, F. Peng, P. Zhang, *Chem Soc Rev* **2019**, *48*, 4178.

[190] J. Mei, Y. W. Zhang, T. Liao, Z. Q. Sun, S. X. Dou, *Natl Sci Rev* **2018**, *5*, 389.

[191] C. Zhao, X. Wang, J. Kong, J. M. Ang, P. S. Lee, Z. Liu, X. Lu, *ACS Appl Mater Interfaces* **2016**, *8*, 2372.

[192] K. Chang, W. X. Chen, *ACS Nano* **2011**, *5*, 4720.

[193] J.-Z. Wang, L. Lu, M. Lotya, J. N. Coleman, S.-L. Chou, H.-K. Liu, A. I. Minett, J. Chen, *Adv. Energy Mater.* **2013**, *3*, 798.

[194] Y. Jing, Z. Zhou, C. R. Cabrera, Z. F. Chen, *J Phys Chem C* **2013**, *117*, 25409.

[195] J. Pang, R. G. Mendes, A. Bachmatiuk, L. Zhao, H. Q. Ta, T. Gemming, H. Liu, Z. Liu, M. H. Rummeli, *Chem Soc Rev* **2019**, *48*, 72.

[196] C. Chen, X. Xie, B. Anasori, A. Sarycheva, T. Makaryan, M. Zhao, P. Urbankowski, L. Miao, J. Jiang, Y. Gogotsi, *Angew Chem Int Ed* **2018**, *57*, 1846.

[197] T. F. Jaramillo, K. P. Jorgensen, J. Bonde, J. H. Nielsen, S. Horch, I. Chorkendorff, *Science* **2007**, *317*, 100.

[198] Y. Li, H. Wang, L. Xie, Y. Liang, G. Hong, H. Dai, *J Am Chem Soc* **2011**, *133*, 7296.

[199] L. Ma, Y. Hu, G. Zhu, R. Chen, T. Chen, H. Lu, Y. Wang, J. Liang, H. Liu, C. Yan, Z. Tie, Z. Jin, J. Liu, *Chem Mater* **2016**, *28*, 5733.

[200] Y. Liu, J. Wu, K. P. Hackenberg, J. Zhang, Y. M. Wang, Y. Yang, K. Keyshar, J. Gu, T. Ogitsu, R. Vajtai, J. Lou, P. M. Ajayan, Brandon C. Wood, B. I. Yakobson, *Nat Energy* **2017**, *2*, 17127.

[201] J. Zhang, T. Wang, P. Liu, Y. Liu, J. Ma, D. Gao, *Electrochim. Acta* **2016**, *217*, 181.

[202] S. H. Yu, Z. Tang, Y. F. Shao, H. W. Dai, H. Y. Wang, J. X. Yan, H. Pan, D. H. C. Chua, *ACS Appl. Energy Mater.* **2019**, *2*, 5799.

[203] M. Gibertini, M. Koperski, A. F. Morpurgo, K. S. Novoselov, *Nat Nanotechnol* **2019**, *14*, 408.

[204] M. Yoshida, Y. Zhang, J. Ye, R. Suzuki, Y. Imai, S. Kimura, A. Fujiwara, Y. Iwasa, *Sci Rep* **2014**, *4*, 7302.

[205] T. Song, Z. Fei, M. Yankowitz, Z. Lin, Q. Jiang, K. Hwangbo, Q. Zhang, B. Sun, T. Taniguchi, K. Watanabe, M. A. McGuire, D. Graf, T. Cao, J. H. Chu, D. H. Cobden, C. R. Dean, D. Xiao, X. Xu, *Nat Mater* **2019**, *18*, 1298.

[206] Y. Yu, F. Yang, X. F. Lu, Y. J. Yan, Y. H. Cho, L. Ma, X. Niu, S. Kim, Y. W. Son, D. Feng, S. Li, S. W. Cheong, X. H. Chen, Y. Zhang, *Nat Nanotechnol* **2015**, *10*, 270.

[207] Z. Wang, Y. Y. Sun, I. Abdelwahab, L. Cao, W. Yu, H. Ju, J. Zhu, W. Fu, L. Chu, H. Xu, K. P. Loh, *ACS Nano* **2018**, *12*, 12619.

[208] D. M. Seo, J.-H. Lee, S. Lee, J. Seo, C. Park, J. Nam, Y. Park, S. Jin, S. Srivastava, M. Kumar, Y. M. Jung, K.-H. Lee, Y.-J. Kim, S. Yoon, Y. L. Kim, P. M. Ajayan, B. K. Gupta, M. G. Hahm, *ACS Photonics* **2019**, *6*, 1379.

[209] Z. Wang, L. Chu, L. Li, M. Yang, J. Wang, G. Eda, K. P. Loh, *Nano Lett* **2019**, *19*, 2840.

[210] Y. Cao, V. Fatemi, A. Demir, S. Fang, S. L. Tomarken, J. Y. Luo, J. D. Sanchez-Yamagishi, K.

Watanabe, T. Taniguchi, E. Kaxiras, R. C. Ashoori, P. Jarillo-Herrero, *Nature* **2018**, *556*, 80.

[211] Y. Cao, V. Fatemi, S. Fang, K. Watanabe, T. Taniguchi, E. Kaxiras, P. Jarillo-Herrero, *Nature* **2018**, *556*, 43.

[212] Y. Chen, Z. Fan, Z. Zhang, W. Niu, C. Li, N. Yang, B. Chen, H. Zhang, *Chem Rev* **2018**, *118*, 6409.

[213] H. Duan, N. Yan, R. Yu, C. R. Chang, G. Zhou, H. S. Hu, H. Rong, Z. Niu, J. Mao, H. Asakura, T. Tanaka, P. J. Dyson, J. Li, Y. Li, *Nat Commun* **2014**, *5*, 3093.

[214] S. Gao, Y. Lin, X. Jiao, Y. Sun, Q. Luo, W. Zhang, D. Li, J. Yang, Y. Xie, *Nature* **2016**, *529*, 68.

[215] A. J. Mannix, X. F. Zhou, B. Kiraly, J. D. Wood, D. Alducin, B. D. Myers, X. Liu, B. L. Fisher, U. Santiago, J. R. Guest, M. J. Yacaman, A. Ponce, A. R. Oganov, M. C. Hersam, N. P. Guisinger, *Science* **2015**, *350*, 1513.

[216] X. Liu, Z. Zhang, L. Wang, B. I. Yakobson, M. C. Hersam, *Nat Mater* **2018**, *17*, 783.

[217] Y. Deng, Y. Yu, Y. Song, J. Zhang, N. Z. Wang, Z. Sun, Y. Yi, Y. Z. Wu, S. Wu, J. Zhu, J. Wang, X. H. Chen, Y. Zhang, *Nature* **2018**, *563*, 94.

[218] N. Briggs, B. Bersch, Y. Wang, J. Jiang, R. J. Koch, N. Nayir, K. Wang, M. Kolmer, W. Ko, A. De La Fuente Duran, S. Subramanian, C. Dong, J. Shallenberger, M. Fu, Q. Zou, Y. W. Chuang, Z. Gai, A. P. Li, A. Bostwick, C. Jozwiak, C. Z. Chang, E. Rotenberg, J. Zhu, A. C. T. van Duin, V. Crespi, J. A. Robinson, *Nat Mater* **2020**, *19*, 637.

[219] Z. Cai, B. Liu, X. Zou, H. M. Cheng, *Chem Rev* **2018**, *118*, 6091.

[220] G. H. Han, N. J. Kybert, C. H. Naylor, B. S. Lee, J. Ping, J. H. Park, J. Kang, S. Y. Lee, Y. H. Lee, R. Agarwal, A. T. Johnson, *Nat Commun* **2015**, *6*, 6128.

[221] H. Gao, J. Suh, M. C. Cao, A. Y. Joe, F. Mujid, K. H. Lee, S. Xie, P. Poddar, J. U. Lee, K. Kang, P. Kim, D. A. Muller, J. Park, *Nano Lett* **2020**, *20*, 4095.

[222] T. Zhang, K. Fujisawa, F. Zhang, M. Liu, M. C. Lucking, R. N. Gontijo, Y. Lei, H. Liu, K. Crust, T. Granzier-Nakajima, H. Terrones, A. L. Elias, M. Terrones, *ACS Nano* **2020**, *14*, 4326.

[223] M. M. Ugeda, A. Pulkin, S. Tang, H. Ryu, Q. Wu, Y. Zhang, D. Wong, Z. Pedramrazi, A. Martin-Recio, Y. Chen, F. Wang, Z. X. Shen, S. K. Mo, O. V. Yazyev, M. F. Crommie, *Nat Commun* **2018**, *9*, 3401.

[224] B. Cho, J. Yoon, S. K. Lim, A. R. Kim, D. H. Kim, S. G. Park, J. D. Kwon, Y. J. Lee, K. H. Lee, B. H. Lee, H. C. Ko, M. G. Hahm, *ACS Appl Mater Interfaces* **2015**, *7*, 16775.

[225] S. Zhang, J. Wang, N. L. Torad, W. Xia, M. A. Aslam, Y. V. Kaneti, Z. Hou, Z. Ding, B. Da, A. Fatehmulla, A. M. Aldhafiri, W. A. Farooq, J. Tang, Y. Bando, Y. Yamauchi, *Small* **2020**, *16*, 1901718.

[226] W. Y. Chen, X. Jiang, S. N. Lai, D. Peroulis, L. Stanciu, *Nat Commun* **2020**, *11*, 1302.

[227] P. T. Loan, W. Zhang, C. T. Lin, K. H. Wei, L. J. Li, C. H. Chen, *Adv Mater* **2014**, *26*, 4838.